\documentclass{aa}
\usepackage{natbib}
\usepackage{graphicx}
\usepackage{amssymb}
\usepackage{multirow}
\usepackage[varg]{txfonts}

\bibpunct{(}{)}{;}{a}{}{,}

\begin{document}

\title{Properties of the circumgalactic medium in simulations compared to observations}

\author{Rubens E.~G.~Machado\inst{1,2,3}
   \and Patricia B.~Tissera\inst{3,4} 
   \and Gast\~ao B.~Lima Neto\inst{5}
   \and Laerte Sodr\'e Jr.\inst{5}}
\institute{Departamento Acad\^emico de F\'isica, Universidade Tecnol\'ogica Federal do Paran\'a, Rua Sete de Setembro 3165, Curitiba, Brazil\\
\email{rubensmachado@utfpr.edu.br}
\and Universidade Federal de Ouro Preto, Departamento de F\'isica, Campus Universit\'ario Morro do Cruzeiro, Ouro Preto, Brazil
\and Departamento de Ciencias F\'isicas, Universidad Andr\'es Bello, Av. Rep\'ublica 220, Santiago, Chile
\and Millennium Institute of Astrophysics, Av. Republica 220, Santiago, Chile
\and Instituto de Astronomia, Geof\'isica e Ci\^encias Atmosf\'ericas, Universidade de S\~ao Paulo, Rua do Mat\~ao 1226, S\~ao Paulo, Brazil
}

\date{Received 09 May 2016 / Accepted 22 October 2017}

%%%%%%%%%%%%%%%%%%%%%%%%%%%%%%%%%%%%%%%%%%%%%%%%%%%%%

\abstract 
%context
{Galaxies are surrounded by extended gaseous halos which store significant fractions of chemical elements. These are syntethized by the stellar populations and later ejected into the circumgalactic medium (CGM) by different mechanism, of which supernova feedback is considered one of the most relevant.}
%aims
{We explore the properties of this metal reservoir surrounding star-forming galaxies in a cosmological context aiming to investigate the chemical loop between galaxies and their CGM, and the ability of the subgrid models to reproduce observational results.}
%methods
{Using cosmological hydrodynamical simulations, we analyse the gas-phase chemical contents of galaxies with stellar masses in the range $10^{9} - 10^{11}\,{\rm M}_{\sun}$. We estimate the fractions of metals stored in the different CGM phases, and the predicted \ion{O}{vi} and \ion{Si}{iii} column densities within the virial radius.}
%results
{We find roughly $10^{7}\,{\rm M}_{\sun}$ of oxygen in the CGM of simulated galaxies having $M_{\star}{\sim}10^{10}\,{\rm M}_{\sun}$, in fair agreement with the lower limits imposed by observations. The $M_{\rm oxy}$ is found to correlate with $M_{\star}$, at odds with current observational trends but in agreement with other numerical results. The estimated profiles of \ion{O}{vi} column density reveal a substantial shortage of that ion, whereas \ion{Si}{iii}, which probes the cool phase, is overpredicted. Nevertheless, the radial dependences of both ions follow the respective observed profiles. The analysis of the relative contributions of both ions from the hot, warm and cool phases suggests that the warm gas ($ 10^5~{\rm K} < T < 10^6~{\rm K}$) should be more abundant in order to bridge the mismatch with the observations, or alternatively, that more metals should be stored in this gas-phase. These discrepancies provide important information to improve the subgrid physics models. Our findings show clearly the importance of tracking more than one chemical element and the difficulty of simultaneously satisfying the observables that trace the circumgalactic gas at different physical conditions. Adittionally, we find that the X-ray coronae around the simulated galaxies have luminosities and temperatures in decent agreement with the available observational estimates.}
%conclusions
{}


\titlerunning{Properties of the CGM in simulations}
\authorrunning{Machado et al.}
\maketitle

%%%%%%%%%%%%%%%%%%%%%%%%%%%%%%%%%%%%%%%%%%%%%%%%%%%%%
\section{Introduction}

% intro
The existence of extended gaseous halos surrounding galaxies has long been a fundamental prediction of galaxy formation theory \citep{WhiteRees1978}, as a natural consequence of gas collapse in the hierarchical build-up of structure. In a sense, the so-called circumgalactic medium (CGM) connects the large-scale intergalactic medium (IGM) with the interstellar medium (ISM). The structure and history of the CGM remain unclear, but substantial progress has been made in recent years, both through observations and simulations \citep[for a review, see][]{Putman2012}.

% feedback in simulations
Cosmological simulations allow for a detailed study of the predicted properties of galaxies and their halos. Early simulations of galaxy formation lacked adequate supernova (SN) feedback and predicted overly dense and massive galaxies. Currently, it is well established that efficient feedback is needed to curtail excessive star formation (SF) and produce realistic disc galaxies satisfying several observed relations \citep[e.g.][and references therein]{Schaye2010, Scannapieco2012, Aumer2013, Vogelsberger2013}. Furthermore, the modelling of SN feedback  including chemical evolution has allowed simulations to describe the enrichment of baryons  as galaxies form and evolve in a cosmological context \cite[e.g.][]{Mosconi2001, Lia2002, Pilkington2012, Tissera2012, Gibson2013}.

% feedback ejects metals
An important consequence of SN feedback is the triggering of galactic-scale winds, which can eject chemical enriched gas from galaxies into the surrounding medium. Therefore the CGM plays a twofold role in the chemical history of a galaxy: it receives enriched material that was expelled in the form of outflows, but it also acts as a reservoir of fuel for future star formation, including the infalling metal-poor IGM gas. For these reasons, the study of CGM properties -- its metallicity in particular -- should hold clues to further constrain simulations models.

% simulations
Cosmological hydrodynamical simulations have been successfully used to study the physical conditions and composition of the CGM \citep[e.g.][]{Oppenheimer2012, Stinson2012, Shen2012, Shen2013, Ford2014, Oppenheimer2016, Suresh2015, Suresh2017, Muratov2017}. With these simulations, it is possible to follow the baryon cycle of galactic inflows and outflows, in order to understand the processes by which galaxies obtain, transform and eject their gas. However, the exact picture remains unclear. The properties of the CGM can be affected by the subgrid physics, the chemical model or the SN feedback scheme adopted, among others. Wind models have different impacts on the CGM. \cite{Marinacci2014} analysed the properties of the CGM in their simulations and found that it is substantially enriched, but the stellar metallicities gradient is found to be too shallow with respect to Galactic observations. Using hydrodynamical simulations, \cite{Cen2013} provided a detailed account of the composition of the cool, warm and hot gas around low-redshift galaxies. \cite{Sokolowska2015} used simulations of Milky Way analogs to study how the ejection of hot gas affects the `missing baryon problem' (referring to the issue that observationally the budget of baryons is still not fully accounted for). Studying low ions and high ions in the CGM of simulated galaxies, \cite{Churchill2015} pointed out that \ion{H}{i} absorption and \ion{O}{vi} absorption arise from physically distinct gas structures, raising the question of how to infer metallicities and total masses. 

% winds
\cite{Suresh2015} explored how the specific wind energies affect the extent to which winds propagate, and found that while higher energy winds produce more extended metal (\ion{C}{vi}) distributions they also generate less stars than expected from the cosmic star formation rate (SFR). \cite{Ford2015} also used simulations with different prescriptions for the wind models, extracting quantities such as equivalent widths and ion ratios to compare with observations. They found that the low-ionization ions (such as \ion{Si}{ii} or \ion{Si}{iii}) were in good agreement, but the high-ionization ones (such as \ion{O}{vi}) were underpredicted. \cite{Oppenheimer2016} analysed low-redshift ($z\sim0.2$) circumgalactic \ion{O}{vi} in simulations from the EAGLE project. Although much of the oxygen is to be found in the CGM, they find that the \ion{O}{vi} column densities are underpredicted with respect to observational constraints.

% observations
Observationally, the CGM is difficult to investigate, mainly due to the low density of the gas, which is spread over great volumes, and also due to the fact that the most interesting lines, like \ion{O}{vi}, have rest-frame wavelengths in the ultraviolet. Even for local galaxies, the picture is still uncertain. The detection of X-ray coronae around spiral galaxies poses significant observational difficulties, because the diffuse and extended nature of the gas leads to low signal-to-noise ratios, even for massive galaxies, which should have the brightest X-ray emission. Nevertheless, there are many instances where hot gas ($T > 10^6$ K) has been inferred in this manner \citep[e.g.][]{Strickland2004, Tullmann2006, Anderson2011, Dai2012, Bogdan2013, Li2013a}, with typical temperatures of $kT \sim 0.6-0.7$ keV for $M_{\star} \sim 10^{11}\,{\rm M}_{\sun}$ galaxies. Extended X-ray emission is also routinely observed around giant elliptical galaxies \citep[e.g.][]{OSullivan2001, Mathews2003, Bogdan2011, Goulding2016}. From the theoretical point of view, the existence of extended X-ray coronae had been a long-standing expectation, and the predicted properties of the X-ray emission have been studied via cosmological simulations as well \cite[e.g.][]{Crain2010, Crain2013}. Recently, \cite{Bogdan2015} compared constraints from X-ray observations with those from simulated X-ray luminosities, indicating that the expectations from simulated spiral galaxies are broadly consistent with the available X-ray estimates. However, for giant ellipticals Illustris produces under-luminous X-ray coronae.

% QSO absorption
One effective way of exploring the CGM -- and particularly of evaluating its metal content -- has been the technique of measuring absorption lines in the direction of background sources, usually quasi-stellar objects (QSOs) whose sightlines pierce the halos of galaxies. This type of observation often focuses on \ion{O}{vi} absorption in the halos of low-redshift galaxies \citep[e.g.][]{Prochaska2011}.

% COS survey
In this respect, the ``COS-Halos'' survey \citep{Werk2013,Tumlinson2013,Prochaska2017} is perhaps the best current source of observational data on CGM metals. This survey explores  the circumgalactic gas of low-redshift galaxies using background QSOs whose sightlines pass within 150 kpc of the galaxies. The QSO spectra are observed with the Hubble Space Telescope's Cosmic Origins Spectrograph in the ultraviolet. Some of the key findings were reported in \cite{Tumlinson2011}, namely that the CGM of $\sim L^{*}$ star-forming galaxies holds at least $\sim10^{7}\,{\rm M}_{\sun}$ of oxygen. This suggests that the CGM of massive local galaxies acts as a major reservoir of metals. A comprehensive budget of metals in and around galaxies is found in \cite{Peeples2014}; and see also \cite{Werk2014} and \cite{Zahid2012}. COS-Halos data have generally been employed for several studies of oxygen absorption in the low-redshift Universe \cite[e.g.][]{Fox2013,Stocke2013}. Recently, \cite{Lehner2015} studied the halo of M31, for which multiple QSO sightlines are available, and found an extended and massive CGM. They estimated a metal mass of at least $2\times 10^{6}\,{\rm M}_{\sun}$ within $0.2r_{\rm vir}$, but possibly one order of magnitude larger within the virial radius. 

% high-redshift
For high-redshift galaxies, observational studies of the structure and metal content of the CGM are more challenging \cite[see e.g.][]{Steidel2010, Turner2014, Turner2015}. In particular, at high redshift, the \ion{O}{vi} doublet may fall within the dense Ly$\alpha$ and Ly$\beta$ forests, making the comparison to low-redshift estimates difficult. \cite{Lehner2013} probed the CGM around galaxies at $z \lesssim 1$, and \cite{Lehner2014} were able to measure column densities for \ion{O}{vi} in higher-redshift galaxies, in the approximate range $2 < z \lesssim 3.5$. They conclude that at redshifts $z =2-3$, a substantial fraction of metals has been already ejected by outflows.

% aim
In this paper we aim to analyse the properties of the gas around galaxies in cosmological simulations and in relation to the metals locked into the galaxies with the aim at understading the feedback loop between the galaxies and their CGMs. One should bear in mind that we are analysing simulations that were not specifically tuned to fit any observed property of the CGM. Therefore these comparisons aim to serve both as a further check of the consistency of the simulations, which can help discriminate between models, as well as a contribution to undestanding observational results. We will use simulations performed with a physically motivated supernova (SN) feedback model which is able to generate metal-loaded outflows without the need to introduce mass or size-dependent parameters. 
% Several other properties of the analysed galaxies have been studied in previous papers \citep{Pedrosa2014, Pedrosa2015, Tissera2016} and are analysed in Section \ref{galaxies}.

% sections
This paper is organized as follows. Section 2 describes the simulations and the simulated galaxy catalogue. In Section 3 our results are presented and compared to the available observational data. We discuss and summarize our findings in Section 4.

%--------------------------------------------------------------------
\begin{figure*}
\includegraphics[width=\textwidth]{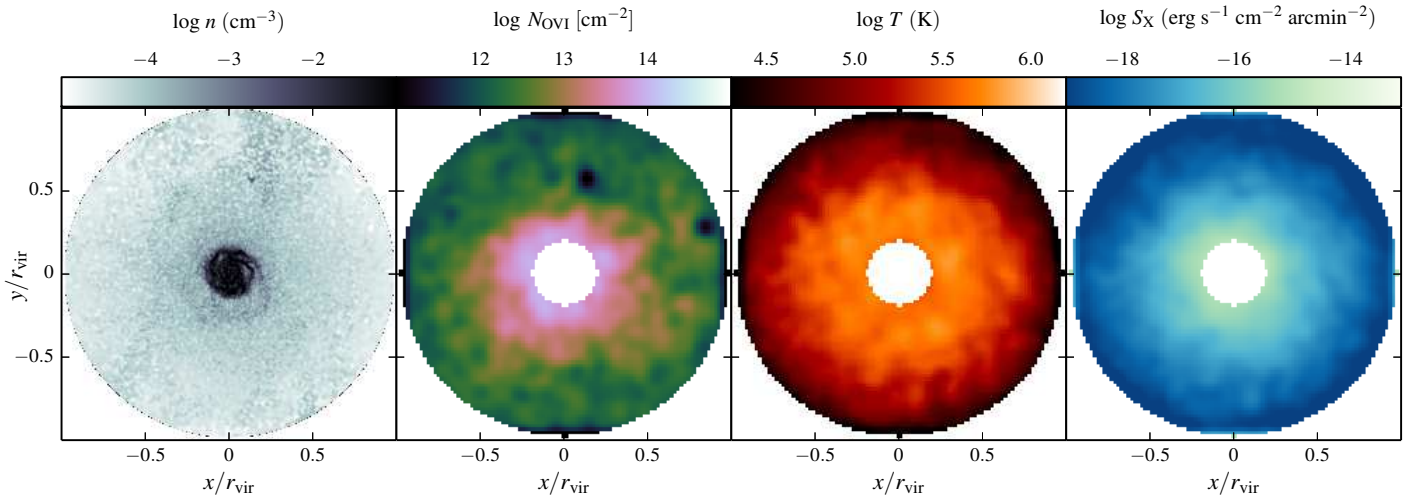}
\caption[]{Projected gas density, \ion{O}{vi} column density, projected temperature, and X-ray surface brightness (from left to right) within the virial radius of a galaxy at $z=0$, from simulation S230D. In the \ion{O}{vi}, temperature and X-ray panels, only the region $r > 2r_{\rm opt}$ is displayed and a gaussian smoothing is applied. This example galaxy has $M_{\star} = 3.6 \times 10^{10} \, {\rm M}_{\sun}$, $r_{\rm vir} = 181 \, {\rm kpc}$ and $r_{\rm opt} = 17 \, {\rm kpc}$ (see Section \ref{galaxies} for definitions).}
\label{maps}
\end{figure*}
%--------------------------------------------------------------------

%%%%%%%%%%%%%%%%%%%%%%%%%%%%%%%%%%%%%%%%%%%%%%%%%%%%%
\section{Numerical Methods}

\subsection{Simulation description} \label{setup}

% cosmology
In this paper we analyse two cosmological simulations from the Fenix project: simulations S230A \citep{DeRossi2010, DeRossi2012, DeRossi2013} and S230D \citep{Pedrosa2015}. Their initial conditions represent a typical field volume of a $\Lambda$CDM universe with the following cosmological parameters: $\Omega_{\rm m} = 0.3$, $\Omega_{\Lambda} = 0.7$,  $\Omega_{\rm b} = 0.045$, a normalization of the power spectrum of $\sigma_{8}= 0.9$ and $H_{0}=100\,h~{\rm km\,s}^{-1}\,{\rm Mpc}^{-1}$ with $h = 0.7$. The simulation volume is a cubic box of comoving side 10 Mpc$\,h^{-1}$, containing $2 \times 230^{3}$ particles. This results in an initial mass resolution of $9.1 \times 10^{5}\,{\rm M}_{\sun}\,h^{-1}$ for the gas particles, and $5.9 \times 10^{6}\,{\rm M}_{\sun}\,h^{-1}$ for the dark matter particles.

% code
The simulations were carried out using the {\sc Gadget-3} code \citep{SpringelHernquist2003, Springel2005}. This version of the code implements a multiphase model for the interstellar medium,  stochastic star formation and metal-dependent radiative cooling. The SN feedback scheme of \cite{Scannapieco2006} is adopted in these runs. It takes into account the effects of both Type II and Type Ia SNe (hereafter, SNII and SNI). Each SN event releases $0.7 \times 10^{51}$ erg of thermal energy into the ISM in the case of S230D; and $1 \times 10^{51}$ erg in the case of S230A. For each star particle where SN events are triggered, the `cold phase' is defined as the gas with temperature $T < 8 \times 10^4$ K and density $\rho > 0.1 \rho_{\rm SF}$ where $\rho_{\rm SF}~=~7 \times 10^{-26}\,{\rm g\,cm}^{-3}$ (or $\sim$ 0.07 cm$^{-3}$) is the threshold of star formation; the `hot phase' is the complement \citep{Scannapieco2005,Scannapieco2006}.

The energy released by the SN explosions is managed differently in the cold and hot phases. The neighbouring hot gas particles thermalize the pumped energy immediately, whereas the neighbouring cold gas particles store the energy in a reservoir for a period of time, until they accumulate sufficient energy to join their nearby hot phase. The decisions are made on particle-particle basis. So that each one regulates the SN feedback effects according to its thermodynamical properties and those of its surrounding medium. \citet{Scannapieco2006} showed that this SN scheme is successfull at driving powerful outflows, which can transport enriched material outside the galaxies. It has been shown that the strength of the SN outflows naturally correlates with the gravitational potential well of the system. Hence, this SN scheme provides a suitable tool to study galaxy formation and metal enrichment in a cosmological context where systems of different masses coexist at a given redshift.

% chemical
The metal ejection and distribution are simultaneous with the explosion for both phases. For SNII, instantaneous recycling is assumed\footnote{\citet{Scannapieco2006} showed that there are not significant differences in the mean metallicities if the time-life from \citet{Raiteri1996} are used  at least for these numerical resolution levels.}. For SNIa, the lifetimes of the progenitors are taken randomly from  a range  $0.1-1$ Gyr (see \citealt{Tissera2016} for a more detailed discussion on this). Initially all baryons are in the form of gas with primordial abundance ($X=0.76$, $Y=0.24$).  A Salpeter initial mass function with minimum and maximum masses of $0.1\,{\rm M}_{\sun}$ and $40\,{\rm M}_{\sun}$ is assumed. We follow twelve chemical species: H, ${}^{4}$He, ${}^{12}$C, ${}^{16}$O, ${}^{24}$Mg, ${}^{28}$Si, ${}^{56}$Fe, ${}^{14}$N, ${}^{20}$Ne, ${}^{32}$S, ${}^{40}$Ca and ${}^{62}$Zn. The nucleosynthesis yields for SNII and SNIa are taken from \cite{Woosley1995} and \cite{Iwamoto1999}, respectively. 

% epsilons
The distribution of energy and chemical SN feedback on the surrounding gas is governed by two free parameters, regulating how much energy and metals are deposited into each phase. These parameters are: $\epsilon_{\rm cold}$, which is the fraction of \textit{energy} injected into the cold phase, and $\epsilon_{\rm met}$, the fraction of \textit{metals} injected into the cold phase. These two parameters can be set independently of each other. The leftover  energy and  metals are dumped into the hot phase.

This chemical model has been shown to reproduce global chemical trends of stellar populations in Milky Way-mass galaxies \citep{Tissera2012} such as the mean high $\alpha$-enhancement and mean old ages of the stars in the bulge compared to those formed in the disc components. Also, \citet{tissera2013} and \citet{tissera2014} discussed the mean abundances of the stellar halos finding similar trends to those observed in nearby galaxies. These results showed that the SN feedback model is able to produce stellar populations with global abundances within observed ranges at $z = 0$. Complementarily \citet{Jimenez2015} studied the formation of a bulge in isolated galaxies reporting that this SN feedback model was able to produce bulges with stellar populations exhibiting the observed SN rates and $\alpha$-enrichment (in the following section we discuss previous results of the simulations we will use in this paper).

% two models
In this work, we analyse two simulations run with the same initial conditions but different SF and SN parameters. The so-called S230D will be taken as the reference simulations based on previous results (see next section). Simulation S230A will be used to assess the effects of varying the parameters mentioned before. Simulation S230A has $\epsilon_{\rm cold}=0.5$ and $\epsilon_{\rm met}=0.5$, while simulation S230D has $\epsilon_{\rm cold}=0.5$ and $\epsilon_{\rm met}=0.8$. The density threshold for star formation in S230D is an order of magnitude larger than in S230A ($\rho_{\rm SF}$ and $0.1 \rho_{\rm SF}$, respectively), while the SN energy released by event is larger in S230A, as mentioned before. The differences in these parameters produce a different regulation of the star formation activity in the dark matter halos, which will have an impact on the CGM properties, as it will be discussed. Both parameters were required to be changed to achieve a better description of chemical abundances. Increasing the star formation threshold decreases the number of stars at high redshift, allowing the formation of more stars at lower redshifts (with the possibility to have more gas to form discs before being strongly consumed by stars, as discussed by \citealt{Tissera2005}). Pumping more metals in the cold phase allows for more suitable levels of enrichment in the simulated discs \citep{Tissera2016} and for a better description of the chemical loop between baryons, i.e. the interplay of the enrichment of gas and stars via SN feedback and star formation, as we will discuss later. We note that feedback from active galactic nuclei (AGNs) is not included in these simulations. However because of the small cosmological volume there are no galaxies massive enough to be affected by this process. % \cite{DeRossi2013}  showed that the mass growth of the halos in the simulated volume is in agreement with the trend measured in the Millennium Simulation.

%%%%%%%%%%%%%%%%%%%%%%%%%%%%%%%%%%%%%%%%%%%%%%%%%%%%%
\subsection{The simulated galaxies} \label{galaxies}

% number of galaxies
We analysed the virialized halos at redshifts $z = 0, 1~{\rm and}~2$, identifying them with a friends-of-friends algorithm. At each redshift, we selected halos with central galaxies more massive than $M_{\star} > 10^{9}\,{\rm M}_{\sun}$ and with at least 2000 gas particles within their virial radius $r_{\rm vir}$ to diminish resolution problems\footnote{The virial radius is the radius at which the mass overdensity is 200 times the  critical density.}. With these cuts, we obtain 82, 73 and 43 galaxies at redshifts $z = 0, 1~{\rm and}~2$ respectively, in simulation S230D. Since we are going to compare our results with the observational sample of \cite{Tumlinson2013}, we estimated the specific SFR of our galaxies, finding that they cover a similar range.

% illustration
As an illustrative example, Fig.~\ref{maps} displays some CGM gas properties of a typical galaxy at redshift $z = 0$. Figure~\ref{maps} shows projected gas density, \ion{O}{vi} column density, projected temperature, and X-ray surface brightness. As can be seen from this figure, the simulated CGM  show  some substructure probably associated with galactic winds and stripped material. The properties of the central galaxies are estimated at the optical radius defined as that which encloses $\sim$80 per cent of the stellar mass. The procedures to estimate \ion{O}{vi} column densities and X-ray luminosities are explained in Sections \ref{column} and \ref{xray}, respectively.

% stellar to virial mass ratio
As mentioned before, we also analysed galaxies in S230A. Because of the lower star formation threshold adopted in this run, gas is  transformed into stars more efficiently at higher redshift. As a consequence, the effects of SN feedback are also stronger at higher redshift, considering that the SN energy released per event is also larger. When halos are less massive, the SN feedback is more efficient at blowing gas away, decreasing the subsequent star formation activity. This leaves fewer galaxies which satisfy the above criteria. The different regulation of star formation produces less massive galaxies in S230A than in S230D.

% previous analyses
The simulated galaxies in S230D and S230A have been analysed in previous works. \citet{Pedrosa2015} found that the disc components in S230D are formed conserving specific angular momentum, and that the bulge components exhibit a correlation between specific angular momentum and stellar mass (or virial mass), in agreement with recent observations. \citet{Tissera2016} analysed the gas-phase metallicity gradients in the discs and the specific star formation rate, finding them to be in agreement with observations at $z = 0$. However, these authors reported galaxies to have steeper gaseous metallicity gradients and lower specific star formation history for higher redshift than reported by observations. This is particularly important for galaxies with stellar masses smaller than $10^{10} {\rm M}_{\sun}$. Hence, the analysis of the simulated CGM will provide clues to understand the origin of this possible discrepancy with observations.

%%%%%%%%%%%%%%%%%%%%%%%%%%%%%%%%%%%%%%%%%%%%%%%%%%%%%
\section{Analysis and Results}

% define CGM
We focus on the analysis of the coronal gas around galaxies, i.e. the gas lying beyond the star-forming disc but within its potential well. The CGM is defined here as the gas contained in the region outside two optical radii but within the virial radius ($2 r_{\rm opt} < r < r_{\rm vir}$) of each galaxy. Substructures, such as satellite galaxies, are found in this region but they are not reckoned as part of the diffuse medium contributing to the  CGM. The gas overdensities above the threshold of star formation ($\rho_{\rm SF} \sim 10^6\,{\rm M}_{\sun}\,{\rm kpc}^{-3}$) are thus removed from the CGM. The ISM is defined as the gas within two optical radii of the central galaxy ($r < 2 r_{\rm opt}$). Whenever we discuss the CGM and the ISM in the analyses of our simulations, these definitions are implied. (But note that in the X-ray analysis of Section~\ref{xray} the overdensities are not removed.)

%--------------------------------------------------------------------
\begin{figure}
\includegraphics[width=\columnwidth]{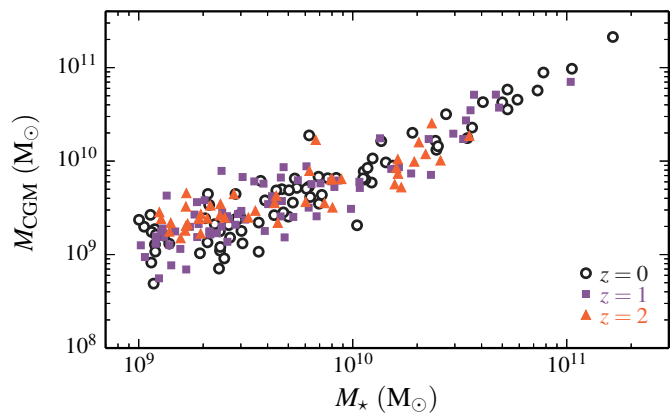}
\caption[]{Total gas mass forming the CGM as a function of stellar mass of the simulated central galaxies at redshifts $z = 0, 1$ and 2 for S230D.}
\label{mass}
\end{figure}
%--------------------------------------------------------------------

Before turning to metal contents and to better characterize a basic feature of the CGM, in Fig.~\ref{mass} we  present their total gas masses as a function of stellar mass, for the three analysed redshifts. The building up of the CGM occurs as star formation proceeds. Galaxies move along the relations with dispersions introduced by their evolutionary histories \citep[e.g.][]{Pedrosa2014}. We can see that the relations between total CGM gas mass and stellar mass are all quite similar to one another for the redshift range $z=0-2$, within the dispersions. We also note that the dispersions are larger for $M_{\star} < 10^{10} {\rm {\rm M}_{\sun}}$.

%--------------------------------------------------------------------
\begin{figure*}
\includegraphics[width=\textwidth]{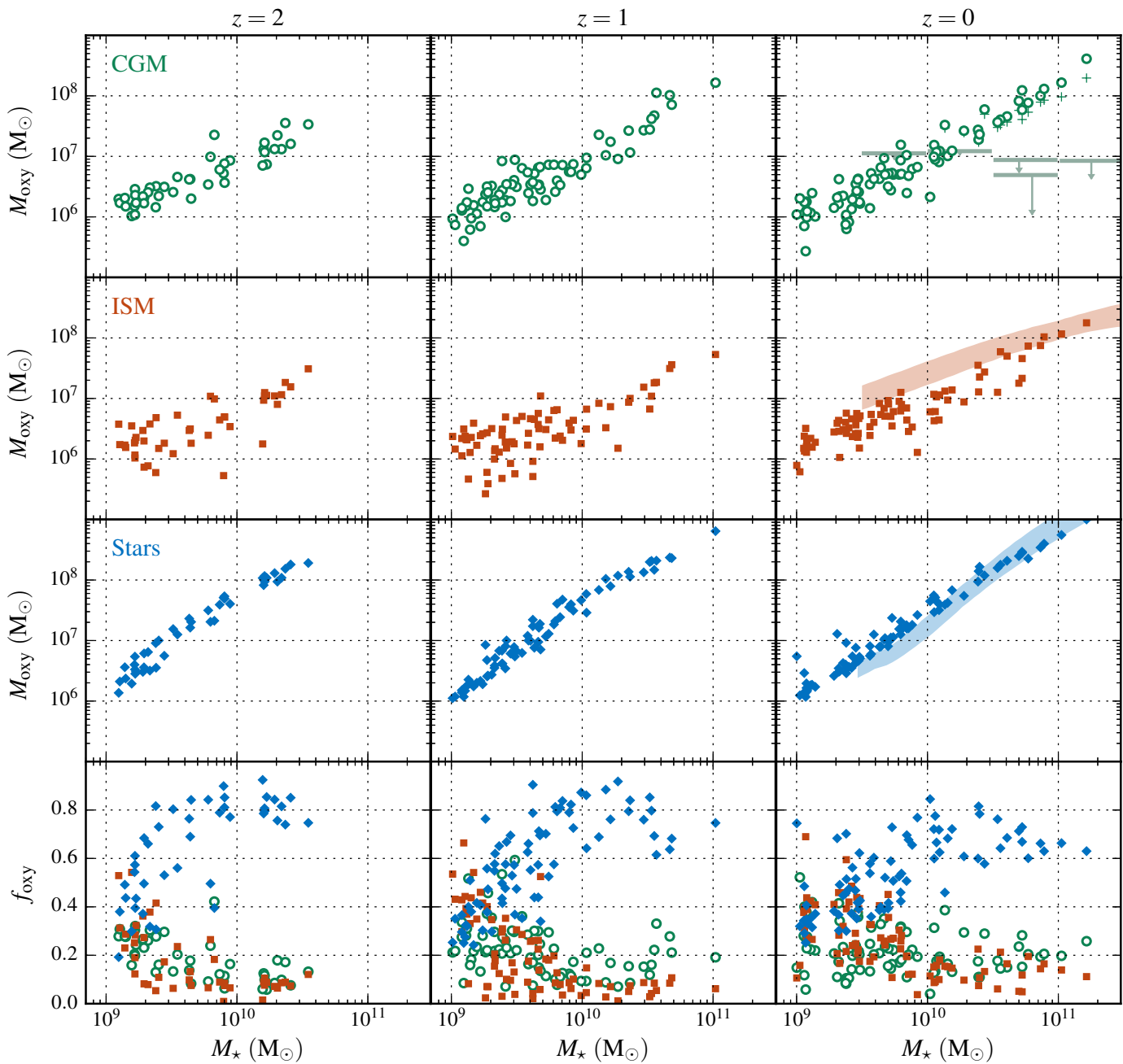}%
\caption[]{Masses of oxygen in the CGM (circles), in the ISM (squares) and in stars (diamonds) as a function of stellar mass, at redshifts $z=2,1$ and 0 (from left to right) for S230D. The observational lower limits of CGM oxygen at $z=0$ are taken from \cite{Tumlinson2011}. The shaded areas for ISM and stars at $z=0$ represent the ranges of observational estimates from \cite{Peeples2014}. The bottom row shows the contribution in each component (CGM, ISM and stars) to the total oxygen mass. In the uppermost right-hand panel, the small crosses represent the estimation of oxygen mass within 150 kpc (shown only for the largest galaxies).}
\label{oxygen3}
\end{figure*}
%--------------------------------------------------------------------

%%%%%%%%%%%%%%%%%%%%%%%%%%%%%%%%%%%%%%%%%%%%%%%%%%%%%
\subsection{Oxygen content}

% COS results
From the observational standpoint, the COS-Halos survey \citep{Tumlinson2013} is one of the best currently available sources of data on the metals of the CGM. Some of the central results about oxygen masses had been reported in \cite{Tumlinson2011}. \cite{Peeples2014} placed those and other results in a broader context by providing a comprehensive budget of metals in and around galaxies, including stars, dust and the gaseous components. They used 28 star-forming galaxies from the COS-Halos sample with redshifts $0.14 < z <0.36$, having stellar masses in the range $\log (M_{\star}/{\rm M}_{\sun})= 9.3-10.8$ with median $\log (M_{\star}/{\rm M}_{\sun})=10.1$ \citep{Werk2012}.

%--------------------------------------------------------------------
\begin{figure*}
\includegraphics[width=0.5\textwidth]{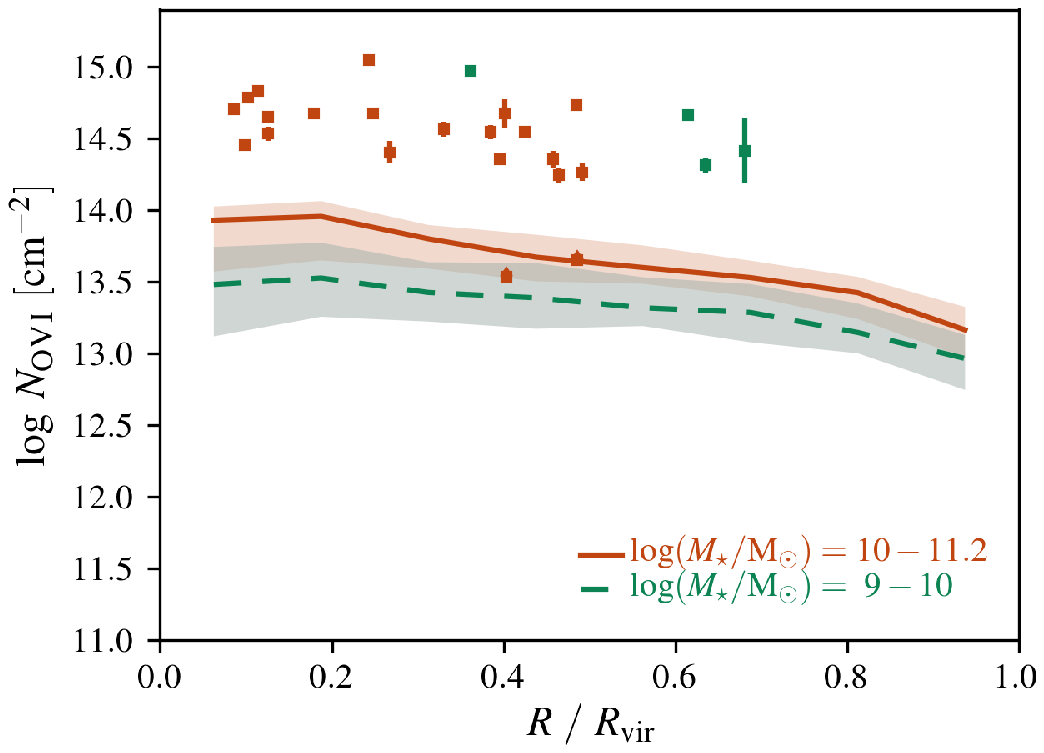}%
\includegraphics[width=0.5\textwidth]{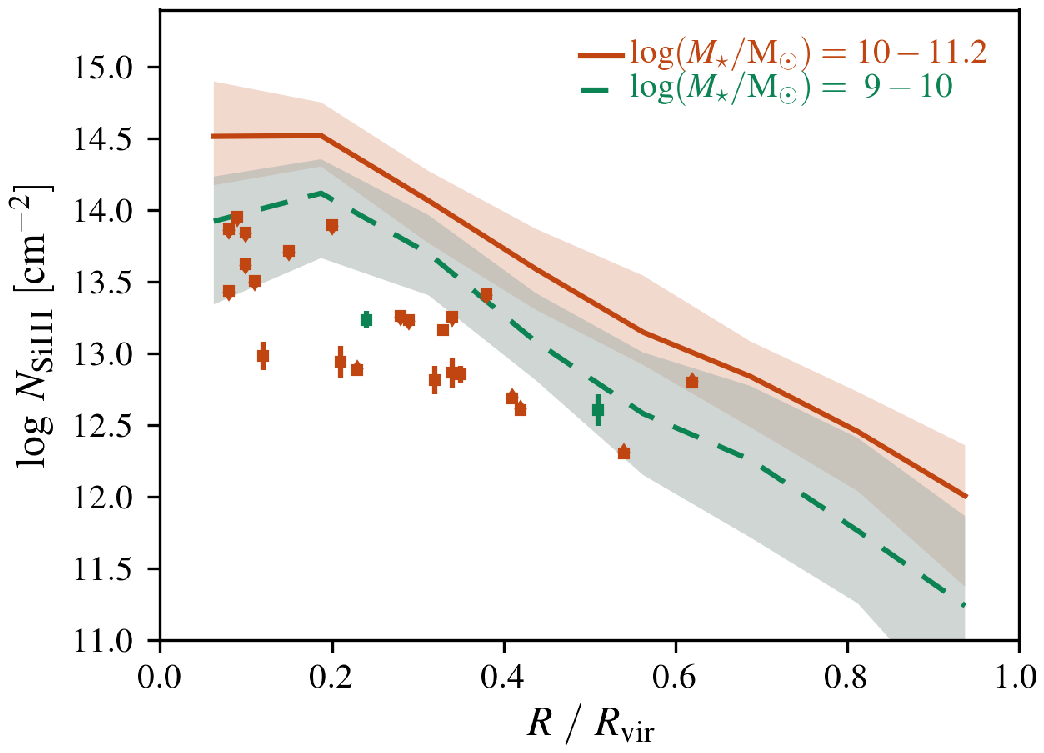}\\
\caption[]{Median profiles of \ion{O}{vi} (left) and \ion{Si}{iii} (right) column density for S230D. The lines show the median values for two mass ranges, and the shaded areas encompass the 16th to 84th percentile range. The points are the observed column densities given by \citet{Tumlinson2011, Tumlinson2013} for the \ion{O}{vi} and by \citet{Werk2014} for the \ion{Si}{iii}. In the observational data only star-forming galaxies are included.}
\label{profilesNoxyb}
\end{figure*}
%--------------------------------------------------------------------

%--------------------------------------------------------------------
\begin{figure*}
\includegraphics[width=0.5\textwidth]{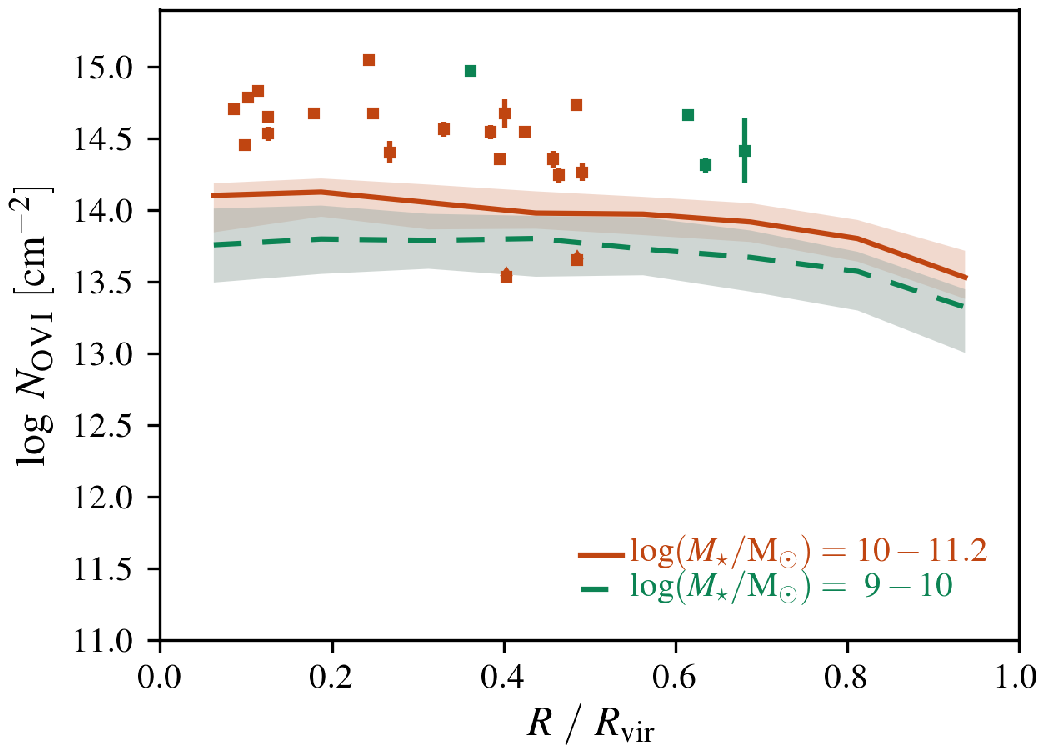}%
\includegraphics[width=0.5\textwidth]{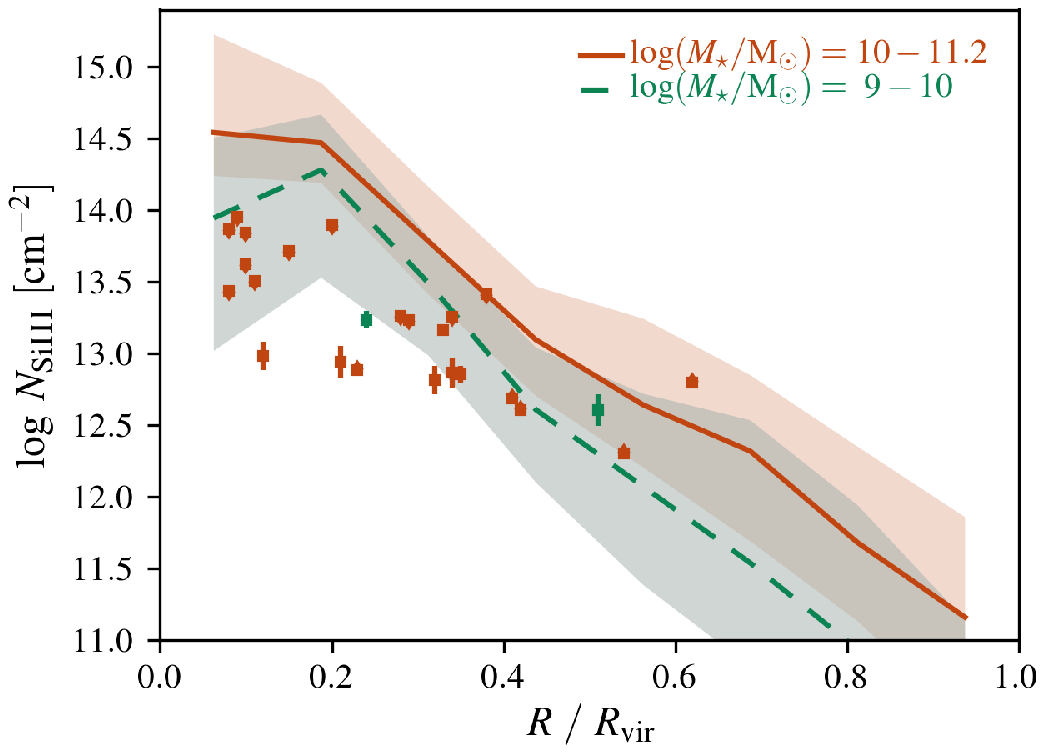}\\
\caption[]{Median profiles of \ion{O}{vi} (left) and \ion{Si}{iii} (right) column density for S230D. Here the ionization fractions were computed with the HM05 UVB model implemented in \textsc{cloudy} version 13.03. Lines and shaded areas analogous to Fig.~\ref{profilesNoxyb}.}
\label{uvb}
\end{figure*}
%--------------------------------------------------------------------

% oxygen mass
In our simulations, we measured the oxygen mass in the stellar population, in the ISM and in the CGM of each galaxy. Figure~\ref{oxygen3} shows them as a function of stellar mass for galaxies in S230D at the different analysed redshifts. We also included the fraction of oxygen mass $f_{\rm oxy}$ held by each component, with respect to the total oxygen mass.

% 150 kpc
In the COS-Halos sample, column densities are measured at impact parameters of as much as 150 kpc. However, considering the virial radii estimated for each of those galaxies, they are being probed only as far out as $\sim$0.7 $r_{\rm vir}$ (cf. Table 2 of \citealt{Tumlinson2013} and Table 1 of \citealt{Werk2014}). In our estimations of the total amount of oxygen in the simulated CGM, we consider the gas contained within the virial radius of each halo, as this can be applied consistently to all mass ranges. In simulation S230D only the 15 most massive objects have virial radii in excess of 150 kpc. In those cases, we also computed the oxygen mass contained within $r<150$ kpc at $z=0$ for comparison. This is shown as the small crosses in the uppermost right-hand panel of Fig.~\ref{oxygen3}. With the exception of the two most massive objects, the differences are far smaller than the scatter of the sample itself. For the low-mass galaxies, extending the CGM definition out to 150 kpc would cause an even more irrelevant change in the total amount of oxygen, because we would be measuring very low-density regions considerably outside the virial radius. In any case, such exercise would be unnecessary, since the  observations we are going to use to compare do not probe impact parameters larger than the virial radius. Indeed, \cite{Suresh2017} point out that even when including large line-of-sight distances, the vast majority of the \ion{O}{vi} absorption takes place within the virial radius. %While taking the virial radius may seem more consistent from a theoretical point of view, the 150-kpc cut does not introduce relevant changes, because these choices differ on whether very low-density gas is taken into account.

The oxygen content of the CGM is shown in the first row of Fig.~\ref{oxygen3}. In the $z=0$ panel we also show the observational data points from \cite{Tumlinson2011}, bearing in mind that they are strictly lower limits. They are computed from the observed column densities by assuming a (maximum) \ion{O}{vi} ionization fraction of 0.2. The actual ionization fraction should be quite lower and thus the total oxygen mass may be considerably higher. In fact, as we will see in the next subsection, given the physical conditions of the simulated CGM gas, the \ion{O}{vi} ionization fraction is found to be at least one order of magnitude lower than 0.2 for most galaxies in our simulations. In the $M_\star \sim 10^{11} {\rm {\rm M}_{\sun}}$ range, the average \ion{O}{vi} ionization fraction is $\sim$0.005. If the CGM gas in the observed galaxies has similar densities and temperatures as in the simulation, then their total amount of CGM oxygen may well be more than two orders of magnitude larger than the lower limit represented by the \cite{Tumlinson2011} points. Indeed, the detailed analysis of the profiles of column densities (Section~\ref{column}) will actually reveal a shortage of \ion{O}{vi} in our simulations.

From Fig.~\ref{oxygen3} we can see that our simulated halos predict a correlation between the $M_{\rm oxy}$ in the CGM and the $M_{\star}$ of the central galaxy for the three analysed redshifts. However, at $z = 0$ current observations do not suggest such a correlation, being consistent with a constant CGM oxygen content across stellar masses. A similar correlation has been reported by \cite{Suresh2017}, with a different simulation code and different subgrid physics. They also find that low-mass galaxies have less oxygen, a trend which is not reflected in observations. In our simulations the $M_{\rm oxy}$ of the CGM, ISM and stars correlates with the $M_{\star}$ of the central galaxies. Except for their zero points, the relations for the ISM and stars are overall consistent with the mass-metallicity relation. The correlation we detect for the CGMs is produced by the regulation of the SN feedback in halos of different masses as we will show below. As a function of redshift, we measured similar relations with  weak evolution. This suggests that simulated systems tend to move along the relations with some dispersion, which might be caused by their different assembly histories \citep{Pedrosa2014}. From these plots, we also conclude that at a given stellar mass, galaxies already have enriched CGMs since $z\sim 2$ \citep{Lehner2013}.

Analysing galaxies from the EAGLE project, \cite{Oppenheimer2016} find that -- for $L^{*}$ galaxies and group halos -- the majority of the oxygen produced and ejected by stars is found beyond the optical extent of the disc, i.e., in the CGM. In our simulations, the total oxygen census shows that the CGM holds roughly as much oxygen as the ISM. However, most of the oxygen of massive galaxies is stored in stars. As can be seen from the bottom row of Fig.~\ref{oxygen3}, the fraction of oxygen stored in stars, $f_{\rm oxy}^{\star}$, increases with stellar mass until  $\sim10^{10}  {\rm {\rm M}_{\sun}}$ and remains at $f_{\rm oxy}^{\star} \sim 0.8$ at $z = 2$. For lower redshift, $f_{\rm oxy}^{\star}$ values are  smaller with an average of $f_{\rm oxy}  \sim 0.6$ by $z = 0$, except for the low-stellar mass galaxies  which  reach values of $f_{\rm oxy}^{\star} \sim 0.3-0.4$. This indicates that a larger fraction of the synthesized metals have failed to be locked into stars. Overall, simulated galaxies with  $M_{\star} < 10^{10} {\rm M_{\sun}}$ show larger variations in their oxygen content, which are close to $f_{\rm oxy}^{\star} \sim 0.4$ on average. As we will see below, this is because of the larger impact of the SN feedback in systems with shallower potential wells \citep{Scannapieco2008}. 

The fraction of oxygen found in the CGM, $f_{\rm oxy}^{\rm \textsc{cgm}}$ shows little evolution since $z=2$. For  galaxies with $M_{\star} < 10^{10} {\rm M_{\sun}}$, which are well sampled in the three analysed redshifts, we find that the percentage of systems with $f_{\rm oxy}^{\rm \textsc{cgm}} > 0.2$ grows from 56 per cent ($z=2$) to 62 per cent ($z=0$), indicating no significant evolution. For more massive galaxies ($M_{\star} > 10^{10} {\rm M_{\sun}}$) which are mostly not yet in place at $z=2$, we would find 38 per cent having $f_{\rm oxy}^{\rm \textsc{cgm}} > 0.2$  at $z=1$ and 35 per cent at $z=0$. However, considering all galaxies the overall level of $f_{\rm oxy}^{\rm \textsc{cgm}}$ is larger at $z=0$, indicating that more metals have been ejected into the CGM. A similar behaviour is present for the ISM. The increasing  fractions of metals in the CGM and ISM of high mass galaxies -- coupled to the fact that metals accumulate over time -- is produced by the fact that these galaxies are able to form more stars and eventually they produce winds strong enough to upload metals into the CGM. In the case of galaxies with $M_{\star} < 10^{10} {\rm M_{\sun}}$, they can more easily drive outflows at a given star formaton rate because of their shallower potential wells.

%--------------------------------------------------------------------
\begin{figure}
\includegraphics[width=\columnwidth]{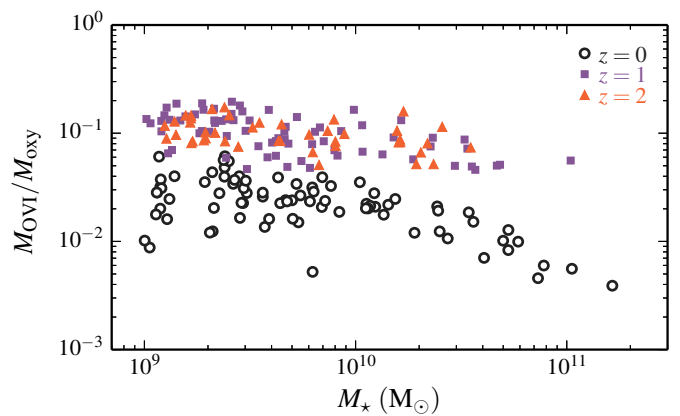}
\caption[]{Ionization fraction of \ion{O}{vi} in the CGM of simulated galaxies in S230D as a function of their stellar mass for the three analysed redshifts.}
\label{massion}
\end{figure}
%--------------------------------------------------------------------

%--------------------------------------------------------------------
\begin{figure*}
\includegraphics[width=0.5\textwidth]{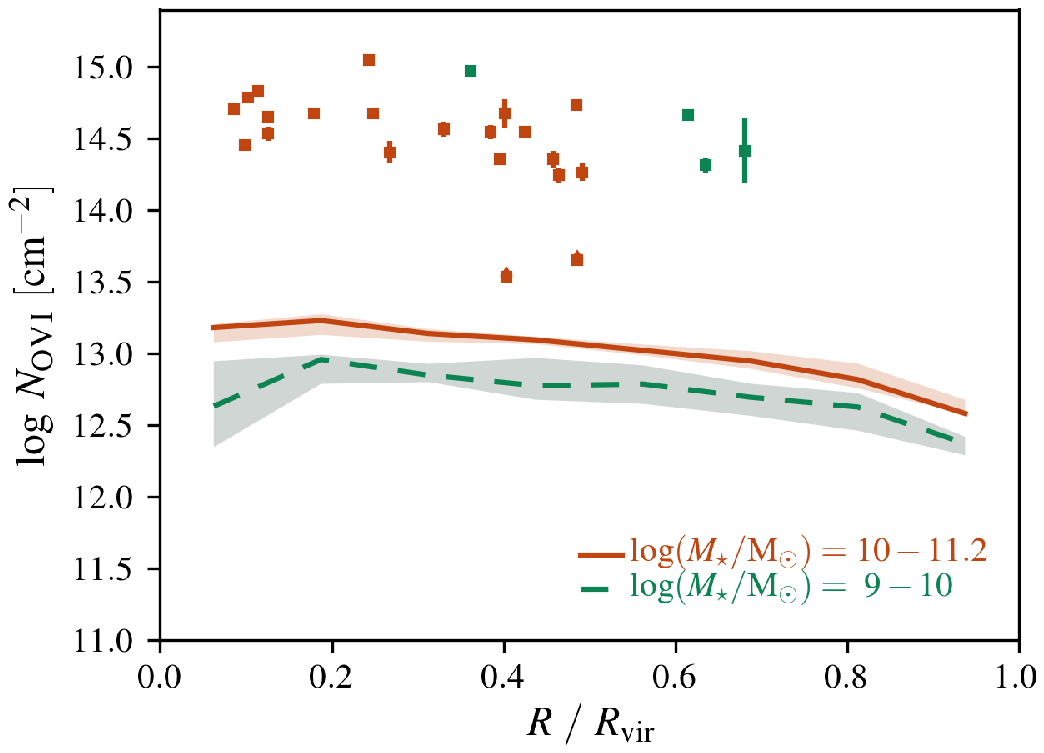}%
\includegraphics[width=0.5\textwidth]{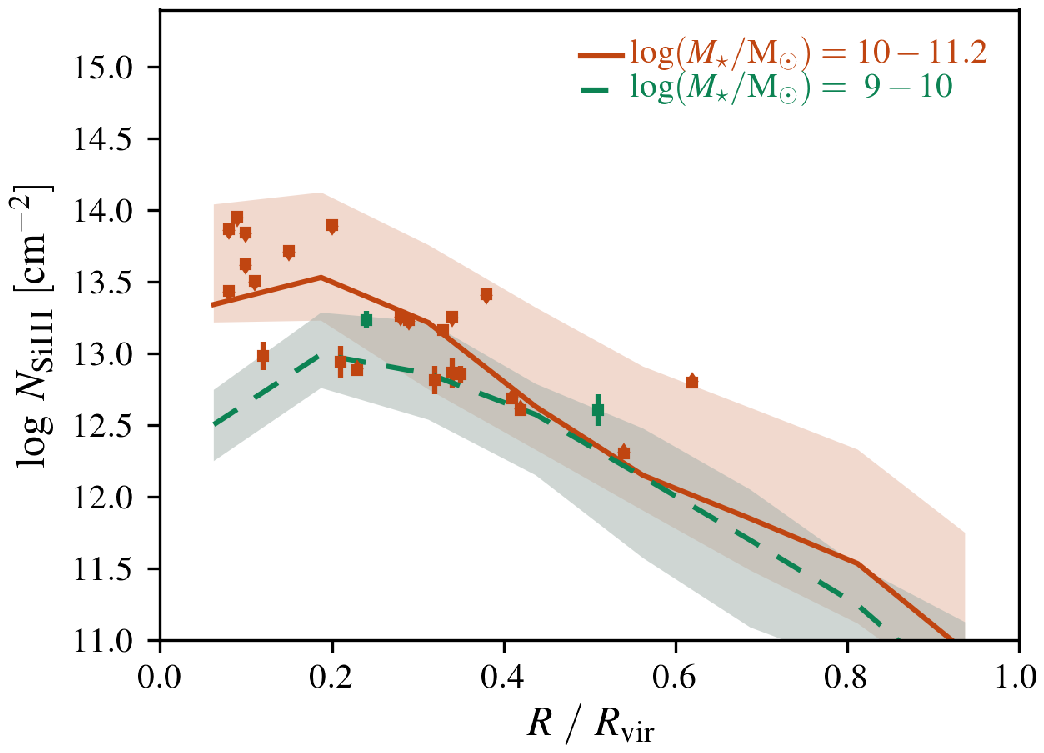}\\
\caption[]{Median profiles of \ion{O}{vi} (left) and \ion{Si}{iii} (right) column density for S230A. Lines and shaded areas analogous to Fig.~\ref{profilesNoxyb}.}
\label{profilesNoxybA}
\end{figure*}
%--------------------------------------------------------------------

%%%%%%%%%%%%%%%%%%%%%%%%%%%%%%%%%%%%%%%%%%%%%%%%%%%%%
\subsection{Column densities} \label{column}

Observations provide information on the \ion{O}{vi} and \ion{Si}{iii} column densities \citep{Tumlinson2011,Tumlinson2013,Werk2013,Werk2014}. These two ions probe mainly the warm and the cool gas respectively (see Section \ref{phases}). In order to derive from our simulations observables that allow for more direct comparisons, we estimate ionization fractions of these species, given the densities and temperatures of the CGM gas.

For this purpose we used \textsc{cloudy} \citep[version 13.03;][]{Ferland2013} in single-zone mode and in the presence of an external radiation background, meaning that both photo-ionization and collisional ionization are taken into account. We adopted the UV background (UVB) of \cite{Faucher2009}, employing the publicly available python scripts\footnote{\texttt{https://github.com/sbird/cloudy\_tables}} of \cite{Bird2015}, which implement self-shielding following \cite{Rahmati2013}. It is then possible to tabulate the desired ionization fractions as a function of density, temperature and redshift, and finally interpolate them at each simulation gas particle. With this information, we can measure radial profiles of column densities as well as median column densities for each galaxy. Thus, the resulting ionization fractions depend not only on the physical conditions of the gas in the simulation (density and temperature), but also on the  adopted UVB. 

% profiles
The profiles of column density as a function of radius are shown in Fig.~\ref{profilesNoxyb}, and they constitute the most direct observables derived from our simulations. The galaxies of simulation S230D at redshift $z=0$ are separated into two mass ranges : $\log(M_{\star}/{\rm M}_{\sun})=$ [9--10] and [10--11.2], and their radial profiles are averaged and show as the lines in Fig.~\ref{profilesNoxyb}, with the shaded areas representing the dispersions. Note that the profiles are expressed normalized to the virial radii. The data points are the observed columns densities from \citet[][only the star-forming galaxies]{Tumlinson2011}. The virial radii for normalization of the observations was taken from \citet{Tumlinson2013}. In our simulations, we do not have massive quiescent galaxies. If we were to apply the same criterium used in \cite{Tumlinson2011} to separate star-forming from quiescent galaxies, namely a cut at ${\rm sSFR} = 10^{-11} {\rm yr^{-1}}$, we would merely remove 14 of the lowest-mass galaxies that happen to have low specific star-formation rates. As this has only a negligible effect in the median lines of Fig.~\ref{profilesNoxyb}, they are kept.

% underpredited
As can be seen from this figure, simulation S230D underpredicts the \ion{O}{vi} column densities by over one order of magnitude for the adopted UVB. At the same time, the \ion{Si}{iii} column densities are overpredicted, indicating the difficulty of simultaneously satisfying both constraints. In the case of the radial distribution, the \ion{O}{vi} profiles display relatively weak radial dependence, while the \ion{Si}{iii} radial profiles are steeper, similarly to observations in both cases. Simulations reproduce the radial dependence better than the column densities.

\cite{Oppenheimer2016}, analysing EAGLE galaxies, also find a shortage of \ion{O}{vi}, with column densities underpredicted by a factor of 2--3.5. Their radial profiles of \ion{O}{vi} column density are as flat as ours. In contrast, those of \cite{Suresh2017} show a steeper decline. Overall, our \ion{O}{vi} predictions for galaxies with $\log(M_{\star}/{\rm M}_{\sun})=$ [9--10.5] fall in similar orders of magnitude as theirs. However, their upper mass bins, particularly $\log(M_{\star}/{\rm M}_{\sun})=$ [10.5--11], are in fact consistent with the corresponding mass range from the observations. In agreement with their results, we also find that low-mass galaxies should have smaller \ion{O}{vi} column densities, a trend not seen in the observational data.

% ion fraction
The ionization fractions of \ion{O}{vi} are shown in Fig.~\ref{massion}, i.e. the ratio between total \ion{O}{vi} mass and the total oxygen mass in the CGM, for each galaxy as a function of stellar mass, at the three analysed redshifts. We see that the \ion{O}{vi} ionization fractions are systematically higher for higher redshifts. The main factor at play here is the redshift-dependence of the adopted UVB, since we had already seen (Fig.~\ref{oxygen3}) that at a given stellar mass the measured CGM oxygen content was roughly constant from $z=2$ to $z=0$. At $z=0$, the \ion{O}{vi} fractions are quite low and even decrease in the most massive galaxies.

Compared to observations, our simulation S230D  underpredicts \ion{O}{vi} column density substantially. It also predicts column densities which are roughly independent of radius. If the physical conditions of the observed galaxies -- i.e. density and temperature of the CGM gas, but also their dependence on stellar mass --  are comparable to those of our simulations, then the total CGM oxygen mass in observed galaxies should be expected to increase with stellar mass. In this sense, the CGM observational constraints of Fig.~\ref{oxygen3} should be regarded strictly as the lower limits they represent. 

Finally, we note that these results are to some degree dependent on the choice of the UVB model. For example, the HM05 UVB model implemented in \textsc{cloudy} version 13.03 yields considerably higher \ion{O}{vi} ionization fractions in the temperature range below $10^5$K, at all relevant densities (at $z=0$). Fig.~1 of \cite{Rahmati2016} illustrates such differences when comparing the ion fractions computed in collisional ionization equilibrium and also in photoionization equilibrium. In collisional equilibrium the peak of \ion{O}{vi} production occurs around $\log T({\rm K})\sim5.5$. In photoionization equilibrium, \ion{O}{vi} ionization fractions are still high for temperatures below this peak. This modifies the relative contributions of the different CGM phases to the production of the ions. As a result of producing more \ion{O}{vi} in the cool phase ($T<10^5$\,K), the profiles of \ion{O}{vi} column density would increase with respect to the results presented above. For comparison we include Fig.~\ref{uvb}, where the  ionization fractions were computed in the case of photoionization equilibrium with the HM05 UVB model implemented in \textsc{cloudy} version 13.03. However, the differences of typically 0.25--0.5 dex would be insufficient to reconcile these estimates with the observations. In the case of \ion{Si}{iii}, the resulting differences are likewise not large enough to affect the current picture significantly.

%%%%%%%%%%%%%%%%%%%%%%%%%%%%%%%%%%%%%%%%%%%%%%%%%%%%%
\subsection{Effects of SN feedback}

As mentioned in Section 2 we also analysed a run with the same initial conditions but different SF and SN parameters. Here we will discuss the main features of the alternative model, simulation S230A. In simulation S230A, a lower threshold of star formation is adopted, more metals are injected into the hot phase, and each SN event releases more energy, with respect to S230D. 

The profiles of \ion{O}{vi} and \ion{Si}{iii} column densities for S230A are shown in Fig.~\ref{profilesNoxybA}. In the case of \ion{Si}{iii}, the column densities are indeed smaller than those of S230D and provide a better agreement with the observational data, indicating that there must be smaller amounts of silicon in the cool phase. However, in the case of \ion{O}{vi}, the column densities also decrease with respect to S230D, falling even shorter of the observational range. The total amount of oxygen in the CGM of S230A galaxies is smaller than in S230D galaxies.

\begin{figure}
\begin{center}
\includegraphics[width=0.95\columnwidth]{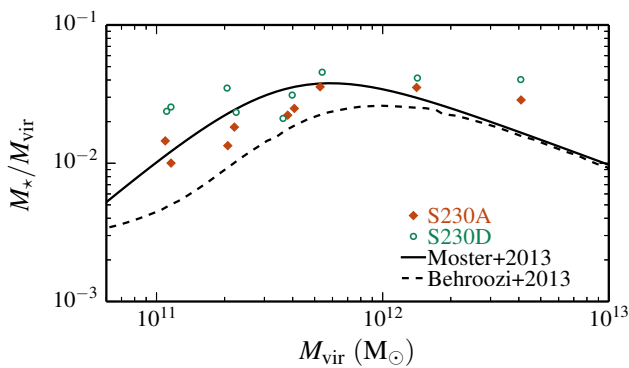}
\end{center}
\caption[]{Ratio of stellar mass to virial mass at $z=0$, as a function of virial mass for the most massive galaxies in simulation S230A (diamonds) and their counterparts in S230D (circles). Also shown are the observationally motivated relations  from \cite{Moster2013} (solid line) and \cite{Behroozi2013} (dashed line).}
\label{mstar3}
\end{figure}

The underprediction is connected to the fact that galaxies in S230A formed fewer stars than their counterparts in S230D. In Fig.~\ref{mstar3} we show the ratio between the stellar mass in the central galaxies and their virial mass as a function of their virial mass for galaxies in both simulations\footnote{We only compare the 9 most massive galaxies in S230A and S230D since they satisfy the selection criteria in both runs at the same time.}. We also included in Fig.~\ref{mstar3} the observationally motivated relations reported by \citet{Behroozi2013} and \citet{Moster2013}. As can be seen, galaxies in S230A have less stellar content than in S230D and seem to follow the observed relations better than those in S230D. However, this fact does not imply that their CGM metals are in better agreement with observations. This discrepancy illustrates the difficulty in reproducing these observational constraints at the same time, and will help us improve the subgrid models.

Specifically, this points to the difficulty of satisfying the abundances of even two chemical elements, i.e. the profiles of column densities of certain ions. Not only are they dependent on the absolute masses of the metals in question, but also on how the local conditions of the CGM affect the ionization fractions of each species, given the adopted UVB.
 
% We also point out that while these simulations satisfy several other properties discussed in the references of Section \ref{galaxies}, they were not tuned to fit CGM parameters.

%--------------------------------------------------------------------
%--------------------------------------------------------------------

%%%%%%%%%%%%%%%%%%%%%%%%%%%%%%%%%%%%%%%%%%%%%%%%%%%%%
\subsection{The CGM phases} \label{phases}

%--------------------------------------------------------------------
\begin{figure*}
\centering
\includegraphics[width=0.9\textwidth]{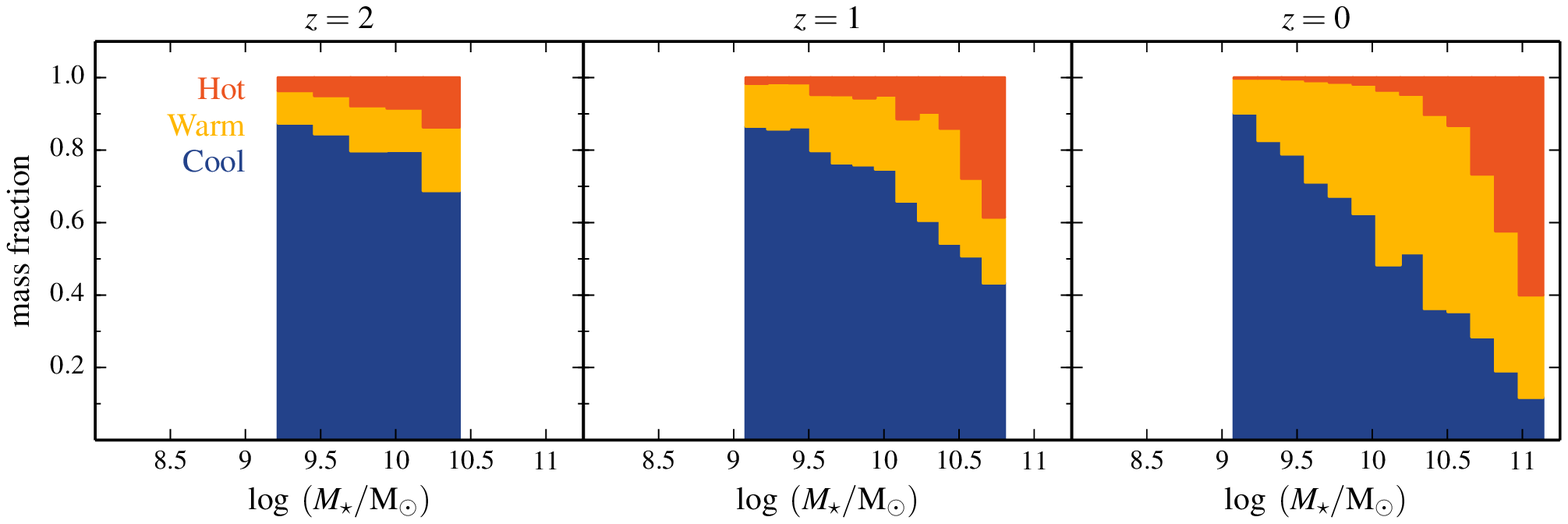}
\includegraphics[width=0.9\textwidth]{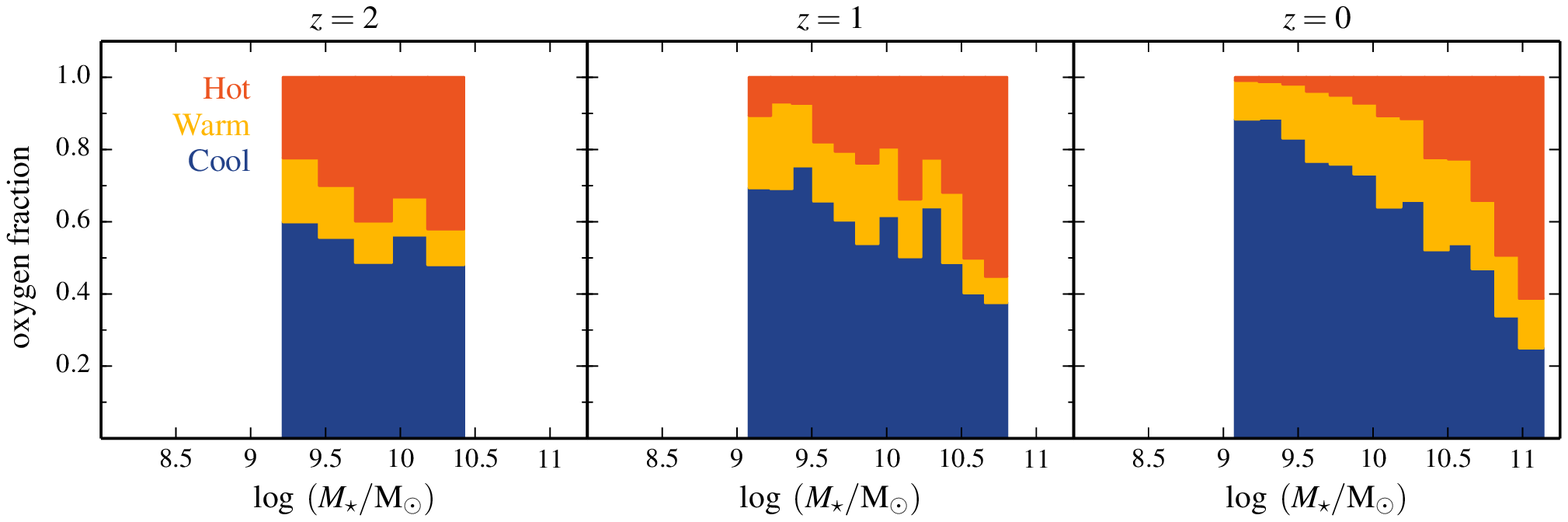}
\includegraphics[width=0.9\textwidth]{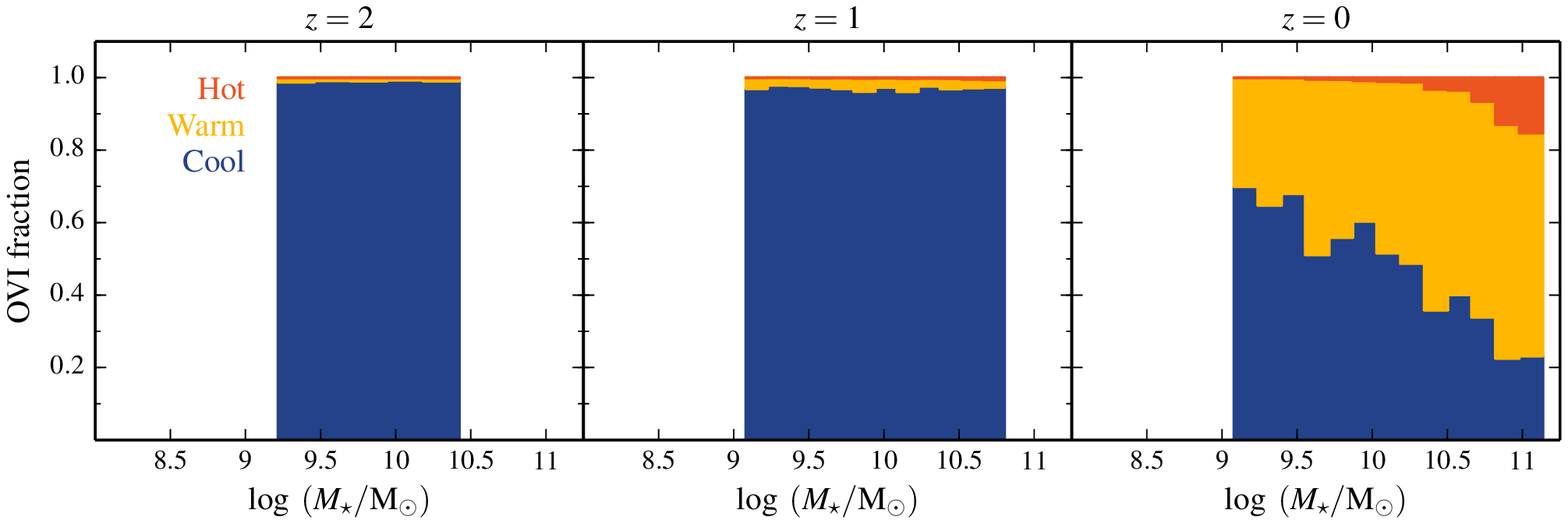}
\caption[]{Fractions of gas mass, oxygen mass, and \ion{O}{vi} mass in each phase of the CGM: cool, warm and hot (in blue, yellow and orange, respectively), as a function of stellar mass, at redshifts, $z=2,1$ and 0 (from left to right) for S230D. The \ion{O}{vi} masses are estimated by adopting the UVB of \cite{Faucher2009}.}
\label{fmass}
\end{figure*}
%--------------------------------------------------------------------

In order to explore the multiphase nature of the CGM gas and how metals are mixed up in these different phases, we adopt the usual convention to define the cool, warm and hot phases:

\begin{itemize}
\item cool: $T < 10^5~{\rm K}$
\item warm: $ 10^5~{\rm K} < T < 10^6~{\rm K}$
\item hot: $T > 10^6~{\rm K}$
\end{itemize}

Figure \ref{fmass} shows how much each phase contributes to the: total CGM mass, the CGM oxygen mass, and the CGM \ion{O}{vi} mass for systems in S230D. The cool phase dominates the total CGM mass at $z=2$, and by $z=1$ the warm/hot gas starts accounting for half the mass in the most massive galaxies. At $z=0$ the massive galaxies ($M_{\star} > 10^{10} {\rm M}_{\sun}$) are dominated by warm/hot gas (64 per cent warm/hot gas on average), while the low-mass ones ($M_{\star} < 10^{10} {\rm M}_{\sun}$) still hold much more cool gas (76 per cent cool gas on average). At $z=0$, the oxygen is distributed across the phases similarly to the total gas, albeit with a relatively smaller contribution of the warm phase. The warm phase is important for the \ion{O}{vi} ion production because, for typical CGM densities of massive galaxies, the \ion{O}{vi} ionization fraction peaks near $T \sim 2 \times 10^5$K and drops rapidly for all other temperatures above and below. Having calculated the ion fractions, we see that at redshift $z=0$ nearly 80 per cent of the \ion{O}{vi} mass comes from the warm/hot phase, in the case of the most massive galaxies. For low-mass galaxies, the \ion{O}{vi} mass is still substantial in the warm phase, but not dominant, due to larger fractions of cool gas. In fact, at redshifts $z=1$ and $z=2$, the UVB yields ionization fractions in the cool temperature range as high as in the warm range; and the availability of cool gas is greater than that of warm gas. As a result, at high redshifts, the vast majority of the \ion{O}{vi} is found in the cool phase.

Regarding silicon, the equivalent of Fig.~\ref{fmass} (not shown) exhibits essentially the same distributions of proportions in the phases for silicon mass as for oxygen mass, indicating merely that the metals are not segregated by phase in the CGM. Specific ions are surely distributed differently depending on the local temperature, and the \ion{Si}{iii} is found very nearly 100 per cent in the cool phase at all redshifts and masses.

Focusing on $z=0$, in order to reach higher \ion{O}{vi} content in the simulations, the warm phase should be much more important than it currently is, assuming the same metal content. One way of achieving this would be to somehow heat the cool CGM gas. Alternatively, if a large amount of the hot CGM gas could be cooled down to below $\sim 10^6$K, it would join the warm phase and thus contribute importantly to the \ion{O}{vi}. On the other hand, the ionization fractions for \ion{Si}{iii} almost vanish above $\sim 10^5$K. Increasing the amount of cool gas at the same metallicity would surely improve the \ion{Si}{iii} column densities, while decreasing the amount of hot gas is irrelevant for \ion{Si}{iii}. So it is not entirely obvious which approach would be better.

\begin{figure}
\includegraphics[width=\columnwidth]{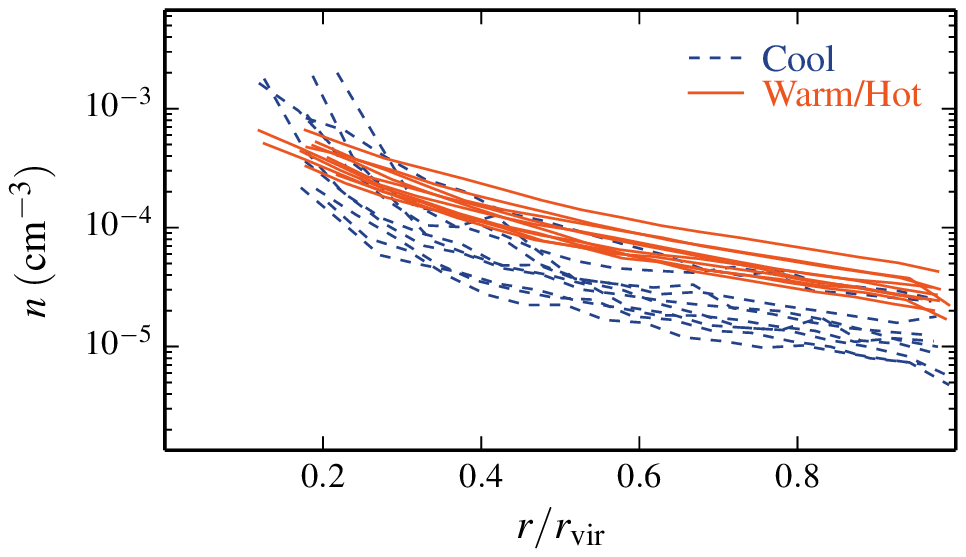}
\caption[]{Density profiles of the CGM gas in the cool phase (dashed lines), and in the warm/hot phase (solid lines). This is a sample of the ten most massive galaxies from S230D at $z=0$.}
\label{profiles}
\end{figure}

% density profiles
Finally, to verify the spatial distribution of the phases, we plot density profiles of the cool phase and of the warm/hot phase in Fig.~\ref{profiles} for the most massive galaxies of S230A at redshift $z=0$. The warm phase is generally more massive throughout the CGM, although the profiles of cool gas are steeper in the inner region, indicating that in this area the cold phase is more massive in some of the haloes (this is because the profiles in Fig.~\ref{profiles} have been measured within spherical shells). Taking the most massive galaxies ($M_{\star} > 10^{10} {\rm M}_{\sun}$), and considering the region $0.4 < r/r_{\rm vir} < 1$, we obtain average indices for the power law fits of $\alpha = -1.81 \pm 0.27$ for the warm/hot phase and $\alpha = -1.93 \pm 0.43$ for the cool phase, consistent with an isothermal profile \citep{Gatto2013}. The inner region ($ r/r_{\rm vir} < 0.4$) of the cool phase has a steeper average profile with $\alpha = -2.52 \pm 0.89$. The radial profiles of \ion{O}{vi} density (not shown) are quite similar in shape to the  density profile of the warm phase in Fig.~\ref{profiles}. The differences in the spatial distribution of the warm/hot and cool phases may explain the different slopes of the column density profiles of \ion{O}{vi} and \ion{Si}{iii} seen in Fig.~\ref{profilesNoxyb}.

%--------------------------------------------------------------------
%--------------------------------------------------------------------

%%%%%%%%%%%%%%%%%%%%%%%%%%%%%%%%%%%%%%%%%%%%%%%%%%%%%
\subsection{X-ray luminosities} \label{xray}

% apec
Given the gas densities, metallicities and temperatures, the X-ray luminosities can be computed from the simulations assuming an appropriate cooling function, $\Lambda(T, Z)$ multiplied by $n^2$, where $n$ is the gas number density. While at high temperature ($T > 10^{7}\,$K) the cooling function is well approximated by bremsstrahlung emission alone ($\propto T^{1/2}$), at lower temperatures one must take into account other processes such as radiative recombination and line emission. These latter processes are particularly sensitive to the gas metallicity $Z$. To estimate the X-ray luminosities, we adopted the APEC model (Astrophysical Plasma Emission Code, \citealt{Smith2001}) implemented on XSPEC~12\footnote{\texttt{http://heasarc.gsfc.nasa.gov/docs/xanadu/xspec}}, that computes the emission spectrum from an optically thin, collisionally-ionized gas. We computed $\Lambda(T, Z)$ in a grid with temperatures in the range $9.2 \times 10^4 < T({\rm K}) < 1.2 \times 10^8$ and the metallicities in the range $0.05 < Z/{\rm Z_{\sun}} < 0.5$, by integrating the emission spectrum in the range [0.01--100 keV]. Following \cite{Crain2010}, we then interpolate this tabulated cooling function for the $(T, Z)$ of each gas particle in the simulation. The integrated $n^2 \Lambda(T, Z)$ over a given volume gives us X-ray luminosities within a given energy range. In principle, all gas particles are taken into account (including the substructures). However, due to their low temperature, only a small fraction of the ISM gas particles are able to contribute in this energy range.

\begin{figure}
\includegraphics[width=\columnwidth]{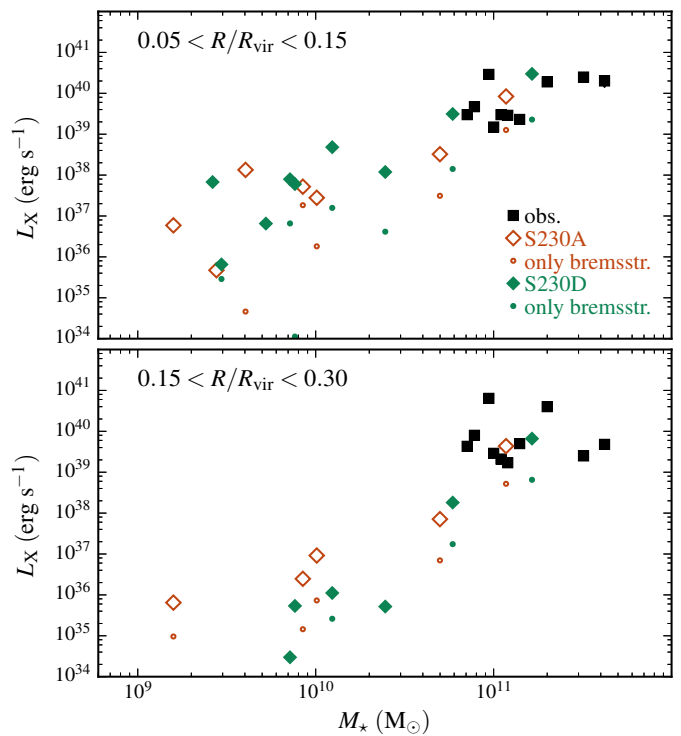}
\caption[]{Simulated X-ray luminosities (0.5--2\, keV) as a function of stellar mass for the most massive galaxies of simulation S230A and their counterparts in S230D, at redshift $z=0$. The diamonds represent luminosities from the simulation, including metal cooling. The squares are the upper-limit observational estimates from \cite{Bogdan2015} in the same energy range. The luminosities (both observed and simulated) were measured within two regions: an inner annulus (upper frame) and an outer annulus (lower frame). For comparison, the small circles show the simulation estimates taking into account only the bremsstrahlung contribution.}
\label{xray2}
\end{figure}

% luminosities
We compare the simulated X-ray luminosities to the results of \cite{Bogdan2015} in the range 0.5--2\, keV. They use \textit{Chandra} observations of eight spiral galaxies, finding no statistically significant X-ray emission beyond the optical disc. Therefore, upper limits are derived. Additionally, their sample is complemented by previous X-ray detections from NGC 1961 and NGC 6753, which are the two most massive galaxies in the sample. They estimate the X-ray luminosities within two circular annuli: an inner one, $0.05 < R/R_{\rm vir} < 0.15$; and an outer one, $0.15 < R/R_{\rm vir} < 0.30$. To allow for a direct comparison, we also measure the X-ray luminosities of our simulated galaxies within the same two regions. In Fig.~\ref{xray2} we display the simulation results (for both S230A and S230D), and we also reproduce the data points from \cite{Bogdan2015}. The simulated X-ray luminosities are computed in the range 0.5--2keV. The observational data are restricted to the high-mass end, where our simulated galaxies are not abundant. Both 230A and 230D predict X-ray luminosities in the range comparable to the observations. In the case of the outer annulus, the agreement is better. In the inner annulus, our simulations may slightly overpredict the X-ray luminosities, bearing in mind that the observational data points are upper limits. However, given the spread of observational values at that mass range, there is an aceptable agreement. The most massive galaxies in S230D have systematically larger X-ray emissions than S230A, but for the lower mass systems this is less clear. The results from the Illustris simulation presented in \cite{Bogdan2015} predict smaller X-ray luminosities. For comparison, the small circles in Fig.~\ref{xray2} show what our X-ray luminosities would be if we took only bremsstrahlung into account. This indicates that in X-ray luminosities, the contribution of enriched warm gas ($10^5 < T < 10^6$\,K) arising from line emissions should be quite relevant in this situation.

% temperatures
As far as the temperatures of the CGM are concerned, it is interesting to compare our results to observations of the X-ray coronae of NGC 1961 and NGC 6753, two galaxies offering a direct estimation of temperature beyond the optical disc \citep{Bogdan2013,Bogdan2015}. The best-fit temperature to the hot gas around these galaxies is $kT\sim0.6$~keV, or $\log T({\rm K})\sim6.8$. If we consider, for example, the four most massive galaxies from S230D and S230A at $z=0$, we find that the luminosity-weighted average temperatures of their CGMs are $\log T({\rm K})\sim 6.8 - 6.9$ (in the 0.5--2keV range). Again, we note that the two observed spirals are slightly more massive than the high-mass end of our simulations; their stellar masses are 4.2 and 3.2 $\times 10^{11}\,{\rm M}_{\sun}$, respectively, whereas our most massive galaxies have $0.8 - 1.6 \times 10^{11}\,{\rm M}_{\sun}$. Therefore, the average CGM temperatures of our most massive galaxies seem to be within the expected values.

\begin{figure}
\includegraphics[width=\columnwidth]{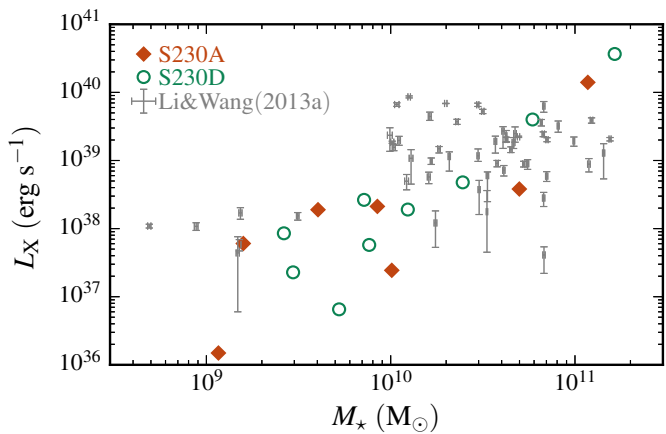}
\caption[]{X-ray luminosities (0.5--2\, keV) as a function of stellar mass, for the most massive galaxies of simulation S230A and their counterparts in S230D, at redshift $z=0$. The points with error bars are the observational results from \cite{Li2013b}.}
\label{LXM}
\end{figure}

Additionally, we compare our simulated X-ray luminosities to the observational data of \cite{Li2013a} in Fig.~\ref{LXM}. In this case, to allow for a direct comparison, the X-ray luminosities were computed in the 0.5--2\,keV range. No density cut is applied; however, the ISM is hardly able to contribute to the diffuse emission in this energy range, since the temperatures of the the dense, star-forming gas ($T\lesssim10^4$\,K) are far lower than 0.5\,keV. Here, where stellar masses are more comparable, the agreement is tolerable although the spread is quite large. In the $10^{10} - 10^{11} \,{\rm M}_{\sun}$ mass range, we see steeper mass dependence of the X-ray luminosities, compared to the observational data of \cite{Li2013a} in that range. There is good agreement between our results (Fig.~\ref{LXM}) and those of \cite{Crain2010} when comparing the mass dependence of the soft X-ray luminosities. It should also be noted that our simulated X-ray luminosities are computed within the virial radius, while those of \cite{Li2013a} come from smaller regions where detection is possible. If we recompute the luminosities within smaller radii we find that the bulk of the emission indeed comes from the inner parts.

%--------------------------------------------------------------------
%--------------------------------------------------------------------

\begin{figure}
\includegraphics[width=\columnwidth]{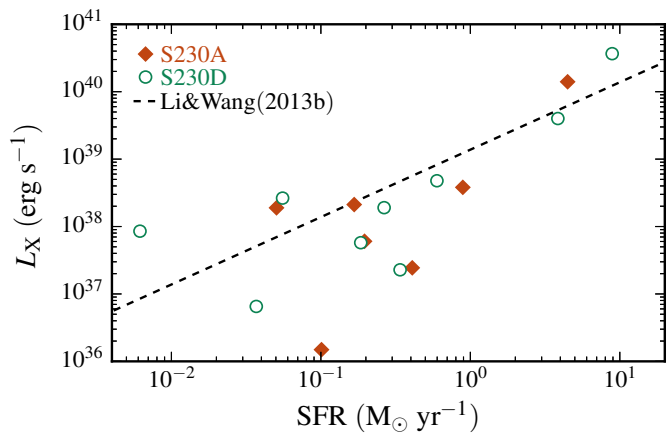}
\caption[]{X-ray luminosities (0.5--2\, keV) as a function of star formation rate, for the most massive galaxies of simulation S230A and S230D, at redshift $z=0$. The dashed line is the observational correlation from \cite{Li2013b}.}
\label{sfr}
\end{figure}

In the present context, coronal X-ray emission should arise from galactic winds and therefore should depend on the SFR \citep{Li2013a, Li2013b, Li2014, Voort2016}. In Fig.~\ref{sfr} we show the correlation between X-ray luminosities in the range 0.5--2\,keV and SFR for our simulations. The dotted line in Fig.~\ref{sfr} is the observational fit from \cite{Li2013b}. Our X-ray luminosities as a function of SFR (Fig.~\ref{sfr}) are similarly in fair agreement with results from \cite{Crain2010}. In the simulations of \cite{Voort2016}, X-ray luminosities are also found to be roughly consistent with observations, although they do not include AGN feedback and neither do we. However, we obtain generally larger $L_{X}$ at a given SFR, making our results more compatible with the \cite{Li2013b} correlation. Although we have fewer objects, the scatter of Fig.~\ref{sfr} is roughly within only one order of magnitude of the observational correlation. \cite{Voort2016} point out that the observations could be biased toward high luminosity and that the simulation lacks the contribution from point sources such as X-ray binaries. Our simulations do not take X-ray point sources into account but offer an acceptable match to the $L_{X}$-SFR correlation. This suggests that galactic winds could be a sufficient mechanism to produce the observed X-ray emission, assuming points sources and AGN feedback can be ignored. As pointed out by \cite{Voort2016}, $L_{X}$ correlates strongly with SFR for galaxies with virial masses below $10^{13}\,{\rm M_{\sun}}$. For more massive galaxies, gas might be heated by accretion shocks in the region of the virial radius, but we have no such objects in our simulation. At least from our most massive galaxies, the $L_{X}$-SFR relation shows no strong dependence on the feedback parameters.

%--------------------------------------------------------------------
%--------------------------------------------------------------------

%%%%%%%%%%%%%%%%%%%%%%%%%%%%%%%%%%%%%%%%%%%%%%%%%%%%%
\subsection{Comparison with previous work}

There are two aspects to the main observational mismatch. One is the shortage of the \ion{O}{vi} column densities, which are underpredicted by an order of magnitude with respect to observations. The other is that in the simulations the oxygen content is always correlated with stellar mass, whereas in the observations there is no dependence. Both issues are also touched upon in recent papers.

In comparison to \cite{Oppenheimer2016}, we find similarly flat profiles of \ion{O}{vi} column density. We are also in agreement about the underprediction of \ion{O}{vi} although the shortage is more pronounced in our results. With regard to the \ion{O}{vi} underprediction, \cite{Oppenheimer2016} offer some possibilities, related to the relative low mass of the simulated galaxies, or to possible Malmquist bias of the COS-Halos sample. Also relevant to our results, they point out that the UVB does not take into account local ionizing sources such as nearby star formation. It is argued that a fossil effect of AGN episodes could enhance \ion{O}{vi} levels by 1 dex under certain conditions. 

Regarding AGN effects, \cite{Segers2017} used simulations to follow the time-variable \ion{O}{vi} abundance in the CGM and foud that even after the AGN fades, \ion{O}{vi} and other ions remain out of ionization equilibrium for several Myr, as a result of delayed recombination. This enhances \ion{O}{vi} column densities by as much as 1 dex within 0.3$R_{\rm vir}$. Interestingly, low ions such as \ion{Si}{iv} are found to decrease (at $z=3$). Both these effects (increasing oxygen and decreasing silicon) hint towards the direction where our simulations lack observational agreement. They also conclude that the the AGN effects are stronger on more massive galaxies, and act predominatly on the cool gas. Even though some of our galaxies are plentiful in cool gas, they might be too low-mass to be affected by AGN feedback. \cite{Mathews2017} also point out that the feedback power needed to maintain a long-lived oxygen-rich CGM may not be easily supplied by galactic supernovae alone, requiring the energy input from AGN.

%--------------------------------------------------------------------
%--------------------------------------------------------------------

In comparison to \cite{Suresh2017}, we find flatter profiles of  \ion{O}{vi} column density and our estimates fall short of the observed range, while their most massive galaxies provide a better match. In agreement with their results, we also find that there should be a mass dependence. The surprising indepedence of \ion{O}{vi} with stellar mass is discussed in \cite{Suresh2017}. This independence would seem to imply that \ion{O}{vi} content must be somehow disconnected from normal scaling properties of the halo. They point out that a strong correlation should be expected. After all, \ion{O}{vi} abundance ultimately depends on mass, metallicity and ionization state of the gas, all of which scale with galaxy mass. Some scenarios were put forth by \cite{Suresh2017}. One is the possibility that smaller galaxies, albeit less enriched, are prone to expel more gas due to their shallower gravitational potentials. However, this would require a fine-tuned balance of the two effects. Furthermore, it should already be accounted for by the simulation. Another possibility would be that intracluster gas is shock-heated to even higher temperatures. This would be relevant to the most massive systems. They also point out that ionization from sources within the galaxy, such as soft X-ray photons could potentially serve to photoionize oxygen at large distances.

Finally, a caveat worth mentioning is the possible presence of small-scale clumps in the CGM that cannot be resolved. Observations suggest that clumps may be as small as 100\,pc \citep{Crighton2015}, which is far below the spatial resolution of simulations. Small cold clouds embedded in warm/hot halo gas could lead to a different temperature structure of the CGM, possibly affecting estimates of average ion fractions and of X-ray emission. Furthermore, as pointed out by \cite{Nelson2016}, such cloud-like structures that give rise to absorption may also have complicated geometries, not even resembling spherical clumps \citep{Churchill2015}.

\section{Conclusions}

In this paper, we analysed the chemical and physical properties of the CGM of simulated galaxies and their evolution since $z = 2$ in a cosmological context. The existence of different gas phases with different chemical abundances is shown and compared to available observations when possible. Complementarily, we studied the fraction of metals stored in the ISM and the stellar populations in order to understand the loop between these components and the CGM surrounding them. 

Our main findings can be summarized as follows:

\begin{enumerate}

\item[(i)]  We measured the total oxygen content of the CGM gas in our simulations and found it to be roughly $\sim10^{7}\,{\rm M}_{\sun}$ of oxygen for galaxies of $M_{\star}\sim10^{10}\,{\rm M}_{\sun}$, in fair agreement with the strictly lower limits imposed by observations. However, the simulated oxygen masses of the CGMs are found to be correlated with the stellar mass of galaxies, a natural outcome of the SN feedback \citep[e.g.][]{Scannapieco2006}, which is at odds with current observations, albeit in agreement with other numerical models \citep{Suresh2017}. In our simulated CGMs, we detect almost no evolution of these relations since redshift $z=2$, suggesting  that galaxies move along the relations as they grow in mass and evolve. Our models predict that at a given stellar mass, CGM metals are already in place by  $z \sim 2$, in agreement with \cite{Lehner2013}.

\item[(ii)] We found that galaxies with $M_{\star} < 10^{9.5}\,{\rm M}_{\sun}$ have a larger fraction of oxygen in the gas-phases than more massive galaxies. This is expected  considering that the SN feedback is more efficient at transporting enriched material outside galaxies with shallower potential wells. Albeit mild, there is a trend for an increase in the fraction of metals in the CGM for massive galaxies as a function of time. The chemical loop existing within the three baryonic components (stars, ISM and CGM), which is set by the adopted SN feedback model, does not allow us to see a global evolution of the metallicity relations. Increasing the SF efficiency to lock the metals into stars is not a solution since the larger fractions of stars produce more SNe, which in turn inject more energy into the ISM blowing aways its gas (Section 3.3). The final outcome is the strong decrease of the SF activity at lower redshift \citep{DeRossi2010}. Because of the physically-motivated SN feedback, we do not find a combination of the parameters which better reproduces the evolution of the mass-metallicity relations \citep[see also ][for similar conclusions with different numerical model]{Suresh2015}.

\item[(iii)] The estimations of the \ion{O}{vi} and the \ion{Si}{iii} offer further interesting conclusions. Once the ionization fractions of \ion{O}{vi} are computed, their column densities reveal a shortage of that ion by $\sim 1$ dex in our simulation at $z=0$. At the same time, the column densities of \ion{Si}{iii} -- an ion that probes the cool phase, contrary to \ion{O}{vi} that probes mainly the warm gas of massive galaxies -- are overpredicted with respect to observations. Note that these simulations have not been deliberately tuned to fit any CGM features, so these discrepancies are useful to improve physical models. Regarding the spatial distributions, the column densities of \ion{O}{vi} and the \ion{Si}{iii} exhibit similar radial dependences to the observations. This reflects the more concentrated distribution of the cool phase compared to the warm/hot phase in the inner regions of the CGM.

\item[(iv)] An alternative subgrid model was explored which injects more metals into the hot phase. It also has a larger SN energy per event and lower star formation threshold. This model does improve the \ion{Si}{iii} column densities, providing a better match with the observations, but here the deficiency of \ion{O}{vi} is even more severe. The different parameters for star formation and for SN feedback also result in more intense star formation activity taking place earlier in the evolution.

\item[(v)] The analysis of the relative contributions of the hot, warm and cool phases reveals that the CGM of low-mass galaxies are mostly dominated by cool gas at redshift $z=0$. The CGM of massive galaxies are dominated by warm/hot gas, but looking at the distribution of oxygen, we see that it is not sufficiently abundant in the warm phase to produce the required amounts of the \ion{O}{vi} ion. In order to obtain larger \ion{O}{vi} content, the warm phase would have to be more prevalent than in the simulations we studied, assuming comparable metallicities. For low-mass galaxies, the contribution of cool gas is possibly too large in our simulated CGM. These results suggest that somehow more heating may be required. Conversely, if it were possible to cool the hot CGM gas of the massive galaxies, that would improve their share of warm gas and thus the \ion{O}{vi} column densities. Not only is it not evident which approach is favoured, but altering the SN feedback parameters introduces various other consequences in the regulation of star formation and the properties of the ISM as well.

\item[(vi)] The predicted X-ray luminosities of the CGM are in the same range as the available observational upper limits. The average CGM temperatures are also in good agreement with observational results. Those observational constraints are derived within annuli extending only as far out as $0.3\,R_{\rm vir}$ for massive spiral galaxies. Again, these simulations were not tuned to fit any X-ray observable. In any case, the column densities of the ions discussed before provide a more stringent constraint, and their scarcity in the simulated CGMs does constitute the most interesting shortcoming which opens the possibility to improve the SN feedback and star formation algorithms. 

\end{enumerate}

%%%%%%%%%%%%%%%%%%%%%%%%%%%%%%%%%%%%%%%%%%%%%%%%%%%%%
\section*{Acknowledgements}

This work  made use of the computing facilities of MareNostrum Barcelona Supercomputer Center, Fenix Cluster (IAFE) and also of the Laboratory of Astroinformatics (IAG/USP, NAT/Unicsul), whose purchase was made possible by the Brazilian agency FAPESP (grant 2009/54006-4) and the INCT-A. We acknowledge support from \textit{Ci\^encia sem Fronteiras} (CNPq, Brazil) and from a FAPESP Visiting Researcher Programme (2012/20038-0-6). The work has been partially supported by Nucleo Milky Way UNAB (Chile), Fondecyt 5113350 (Chile), PICT 959 Mincyt (Argentina) and PIP 2012 Conicet (Argentina). GBLN acknowledges support from CNPq. LSJr also acknowledges support from CNPq and FAPESP grant 2012/00800-4. 

%%%%%%%%%%%%%%%%%%%%%%%%%%%%%%%%%%%%%%%%%%%%%%%%%%%%%
\bibliographystyle{aa.bst}
\bibliography{CGM.bib}

\begin{appendix}

\section{Convergence tests}
 
The SN feedback model adopted to run the analysed simulations is the same one used for the set of Milky-Way mass-like systems of the Aquarius project by \citet{Scannapieco2009}. To study numerical convergence, these authors run two lower resolution versions of one of there halos decreasing the numerical resolution from the order of $10^6$ to $160000$ and $80000$ total particles (the so-called Aq-5, Aq-6 and Aq-7 levels of resolution).  They reported good agreement when testing the convergence of the dynamical properties of baryons \citep{Scannapieco2009,Scannapieco2010} and the dark matter \citep{Tissera2010} distributions. And more relevant to this work, \citet{Tissera2012} analysed the abundances of the stellar populations for the same set of runs finding good convergence for the chemical abundances of the stellar populations identified in the disc, bulge and halos (their figure 5). Since the stars formed from the gas, acquiring its  chemical abundances, they reflect and follow the chemical trends of the gas phase from where they formed. In this sense it is expected they share the same convergence trends. Our analysed halos are resolved with different number of particles: the largest simulated halo has about one million total particles, the lowest one of the order of 10000 total particles. Note also that the mass resolution for our simulations are between the Aq-5 and Aq-6 levels of resolution.

%--------------------------------------------------------------------
\begin{figure}
\centering
\includegraphics[width=0.67\columnwidth]{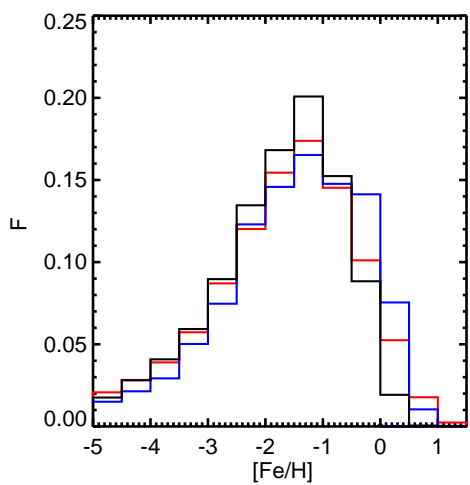}
\caption[]{Distributions of [Fe/H] for the gas particles within the virial radius of the Aq-E set of runs with different level of numerical resolutions, declining from Aq-E-5 (red line), Aq-E-6 (blue line) and Aq-E-7 (black line) (see also \citealt{Scannapieco2009, Scannapieco2011} and \citealt{Tissera2010, Tissera2012}).}
\label{convergence}
\end{figure}
%--------------------------------------------------------------------

\cite{Tissera2012} reported good convergence of the chemical properties of the stellar populations obtained by using the same code adopted for our simulated CGMs (figure 5 of their paper). In order to test if this was also valid for the chemical abundances of the gas components, we extended the analysis to the gas components within the virial radius of the three different resolution runs of the halo Aq-E (5-level, 6-level and 7-level) as part of the Aquarius project (see \citealt{Scannapieco2009, Scannapieco2011} and \citealt{Tissera2010} for details on these simulations). The simulated haloes analysed in this work are within the numerical range covered by these experiments. As shown in Fig.~\ref{convergence}, the [Fe/H] distributions of the gas components in the virial radius agree quite well. We performed Kolmogorov-Smirnoff tests over the distributions finding no significant statistical differences between them (prob.~$\sim$0.9).

%--------------------------------------------------------------------
\begin{figure}
\centering
\includegraphics[width=\columnwidth]{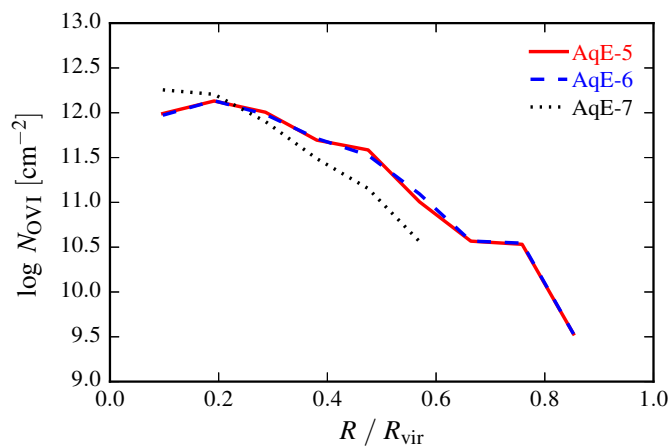}
\caption[]{Convergence test for the \ion{O}{vi} column density, showing Aq-E-5 (red line), Aq-E-6 (blue line) and Aq-E-7 (black line).}
\label{Aq567}
\end{figure}
%--------------------------------------------------------------------

To investigate even further the convergence of the results, we tested the \ion{O}{vi} column density, since this is one of the central results in our analysis. Fig.~\ref{Aq567} shows the radial profiles of \ion{O}{vi} column density for the three levels or resolution of Aq-E. The ion fractions were computed with the same procedures described in Section \ref{column}. This individual Aq-E galaxy happens to have relatively low column densities, compared to the median of the S230D sample, but here we are focused on evaluating the convergence of simulations using the same feedback implementation. From Fig.~\ref{Aq567} one clearly sees that the results are converged above the AqE-6 level. Since the simulations in the present paper have mass resolution intermediate between the levels of Aq-5 and Aq-6, it is expected that they also converged. Finally, we also performed the same resolution test for the profiles of \ion{Si}{iii} column density. We found that the results are very similar to the oxygen resolution test. Since these ions probe different phases of the gas, this test ensures that the observables are well resolved across a wide range of temperatures and densities.

\end{appendix}

\label{lastpage}
\end{document}